# Final results of the measurement to search for rare decays of naturally occurring osmium isotopes with ultra-low background gamma-ray spectrometry

P. Belli[1,2], R. Bernabei[1,2*], F. Cappella[3,4], V. Caracciolo[1,2], R. Cerulli[1,2], F. A. Danevich[1,5], V. Yu. Denisov[5,7,8], A. Incicchitti[3,4], D. V. Kasperovych[5], V. V. Kobychev[5], G. P. Kovtun[9,10,†(Deceased)], M. Laubenstein[6], D. V. Poda[11], O. G. Polischuk[3,5], A. P. Shcherban[9], D.A. Solopikhin[9], S. Tessalina[12], and V. I. Tretyak[5,6]

[1] *INFN, Sezione Roma "Tor Vergata", I-00133 Rome, Italy*

[2] *Dipartimento di Fisica, Università di Roma "Tor Vergata", I-00133 Rome, Italy*

[3] *INFN, Sezione Roma, I-00185 Rome, Italy*

[4] *Dipartimento di Fisica, Università di Roma "La Sapienza", I-00185 Rome, Italy*

[5] *Institute for Nuclear Research of NASU, 03028 Kyiv, Ukraine*

[6] *INFN, Laboratori Nazionali del Gran Sasso, 67100 Assergi (AQ), Italy*

[7] *INFN, Laboratori Nazionali di Legnaro, 35020 Legnaro, Italy*

[8] *Faculty of Physics, Taras Shevchenko National University of Kyiv, 03022 Kyiv, Ukraine*

[9] *National Science Center "Kharkiv Institute of Physics and Technology", 61108 Kharkiv, Ukraine*

[10] *V.N. Karazin Kharkiv National University, 4, 61022 Kharkiv, Ukraine*

[11] *Université Paris-Saclay, CNRS/IN2P3, IJCLab, 91405 Orsay, France*

[12] *John de Laeter Centre for Isotope Research, GPO Box U 1987, Curtin University, Bentley, WA, Australia*

**Abstract.** A long-term measurement was conducted to search for α, double-α and double-β decays with γ quanta emission in naturally occurring osmium isotopes. This study took advantage of two ultra-low background HPGe detectors and one ultra-low background BEGe detector at the Gran Sasso National Laboratory (LNGS) of the INFN. Over almost 5 years of data were taken using high-purity osmium samples of approximately 173 g. The half-life limits set for α decays of $^{184}$Os to the first $2^+$ 103.6 keV excited level of $^{180}$W ($T_{1/2} \geq 9.3 \times 10^{15}$ yr) and of $^{186}$Os to the first $2^+$ 100.1 keV of $^{182}$W ($T_{1/2} \geq 4.8 \times 10^{17}$ yr) exceed substantially the present theoretical predictions that are at level of $T_{1/2} \sim (0.6 - 3) \times 10^{15}$ yr for $^{184}$Os and $T_{1/2} \sim (0.3 - 2) \times 10^{17}$ yr for $^{186}$Os. New half-life limits on the 2EC and EC$\beta^+$ decay of $^{184}$Os to the ground and excited levels of $^{184}$W were set at level of $T_{1/2} > 10^{16} - 10^{17}$ yr; a lower limit on the $2\beta^-$ decay of $^{192}$Os to the $2^+$ 316.5 keV excited level of $^{192}$Pt was estimated as $T_{1/2} \geq 6.1 \times 10^{20}$ yr. The half-life limits for 2α decay of $^{189}$Os and $^{192}$Os were set for the first time at level of $T_{1/2} > 10^{20}$ yr.

* corresponding author, rita.bernabei@roma2.infn.it

## 1. Introduction

The alpha decay continues being a major tool for studying the structure, levels and properties of atomic nuclei to also contribute to developing nuclear models [1, 2, 3]. It is noteworthy that the range of the half-lives ($T_{1/2}$) relative to the α activity spans almost an infinite range, starting from extremely short times on the order of $T_{1/2} \sim 10^{-22} - 10^{-16}$ s for $^{5}$He, $^{5}$Li and $^{8}$Be, to the longest up-to-date observed half-lives $T_{1/2} \sim 10^{18} - 10^{21}$ yr for $^{180}$W, $^{151}$Eu and $^{209}$Bi [4]. However, these long half-lives are not a limit for the α activity's lifetime, they merely reflect the current capability of the contemporary low-counting experimental techniques to detect rare α decays. In this regard, the study of long-lived α decays is of significant interest.

All naturally occurring osmium isotopes are theoretically unstable relative to α decay with an energy release in the range $Q_\alpha \sim 361 - 2959$ keV. However, there has only been an indication of the ground state to ground state α decay of $^{184}$Os, inferred from measurements showing an excess of its progeny $^{180}$W in meteorites and terrestrial rocks with the half-life $T_{1/2} = (1.1 \pm 0.2) \times 10^{13}$ yr [5] and $T_{1/2} = (3.4 \pm 2.1) \times 10^{13}$ yr [6]. These results closely align with the lower limits established by direct laboratory experiments: $T_{1/2} > 2.0 \times 10^{13}$ yr [7] and $T_{1/2} > 5.6 \times 10^{13}$ yr [8]. The tension between the half-life value reported in [5] and the limits obtained in the direct measurements [7, 8] can be explained mainly by underestimation in [5] of the contribution to the measured excess of $^{180}$W from galactic cosmic rays. Indeed, an accurate determination of the contribution led to the higher value and bigger uncertainty of the $^{184}$Os half-life in [6]. Nevertheless, neither the claims nor the limits contradict the theoretical predictions for $^{184}$Os half-life, estimated to be in the range $T_{1/2} = (2.1 - 7.5) \times 10^{13}$ yr [9, 10, 11, 12, 13]. The alpha decay of $^{186}$Os with $T_{1/2} = (2.0 \pm 1.1) \times 10^{15}$ yr was directly observed in an experiment with a semiconductor detector and an enriched $^{186}$Os sample (61.27% enrichment) [14].

Alpha decays to the excited levels of progeny nuclei have not been observed in any of the naturally occurring osmium isotopes. These transitions are accompanied by the emission of γ quanta, which makes it possible to utilize γ-ray spectrometry to search for this α decay. In the case of α decays of $^{189}$Os and $^{192}$Os additional γ quanta are emitted due to the progeny's instability. The energy releases investigated in this study and theoretical predictions for α transitions of naturally occurring osmium isotopes to lower levels of the progeny nuclei (and of $^{189}$Os and $^{192}$Os that have unstable progeny) are given in Table 1. However, the primary objective of the presented measurement was the search for α decays of $^{184}$Os and $^{186}$Os to the first excited levels of their progeny, taking into account the promising theoretical estimations for the half-lives of these processes. Simplified decay schemes of $^{184}$Os and $^{186}$Os are presented in Fig. 1. The decay schemes of other naturally occurring osmium isotopes can be found in [16].

Table 1. The α decay channels of the naturally occurring osmium isotopes studied in the present work (all the transitions are assumed to occur from the ground state of the parent nucleus), decay energy ($Q_\alpha$) [15], isotopic concentration ($\delta$, measured in [16], see Table II in that work), spin and parity ($J^\pi$), energy of the progeny level ($E$), experimental half-life limits obtained in the present study, theoretical predictions for the decay half-lives.

| Transition | $Q_\alpha$ (keV) | $\delta$ (%) | $J^\pi$, $E$ (keV) | Experimental $T_{1/2}$ (yr) | Theoretical $T_{1/2}$ (yr) [9, 10, 11, 12, 13] |
|---|---|---|---|---|---|
| $^{184}Os \to {}^{180}W$ | 2958.8(16) | 0.0170(7) | $2^+$, 103.6 | $\geq 9.3 \times 10^{15}$ | $6.3 \times 10^{14} - 2.9 \times 10^{15}$ |
| | | | $4^+$, 337.6 | $\geq 4.8 \times 10^{16}$ | $9.2 \times 10^{17} - 2.5 \times 10^{19}$ |
| $^{186}Os \to {}^{182}W$ | 2821.2(9) | 1.5908(6) | $2^+$, 100.1 | $\geq 4.8 \times 10^{17}$ | $3.3 \times 10^{16} - 2.2 \times 10^{17}$ |
| | | | $4^+$, 329.4 | $\geq 6.0 \times 10^{18}$ | $7.2 \times 10^{19} - 2.9 \times 10^{21}$ |
| $^{187}Os \to {}^{183}W$ | 2721.7(9) | 1.8794(6) | $3/2^-$, 46.5 | $\geq 3.2 \times 10^{15}$ | $1.6 \times 10^{17} - 4.4 \times 10^{20}$ |
| | | | $5/2^-$, 99.1 | $\geq 1.9 \times 10^{17}$ | $9.1 \times 10^{17} - 2.8 \times 10^{21}$ |
| $^{188}Os \to {}^{184}W$ | 2143.2(9) | 13.253(3) | $2^+$, 111.2 | $\geq 3.3 \times 10^{18}$ | $1.3 \times 10^{28} - 2.9 \times 10^{29}$ |
| | | | $4^+$, 364.1 | $\geq 8.1 \times 10^{19}$ | $8.9 \times 10^{33} - 1.9 \times 10^{36}$ |
| $^{189}Os \to {}^{185}W$ | 1976.2(9) | 16.152(4) | $3/2^-$, g.s. *) | $\geq 3.5 \times 10^{15}$ | $3.1 \times 10^{29} - 2.4 \times 10^{34}$ |
| | | | $1/2^-$, 23.5 | $\geq 3.5 \times 10^{15}$ | $3.2 \times 10^{30} - 1.8 \times 10^{35}$ |
| | | | $5/2^-$, 65.9 | $\geq 7.6 \times 10^{17}$ | $3.1 \times 10^{31} - 2.1 \times 10^{36}$ |
| $^{190}Os \to {}^{186}W$ | 1375.9(12) | 26.250(8) | $2^+$, 122.6 | $\geq 1.2 \times 10^{19}$ | $1.6 \times 10^{51} - 1.1 \times 10^{54}$ |
| | | | $4^+$, 396.5 | $\geq 8.6 \times 10^{19}$ | $1.6 \times 10^{65} - 5.8 \times 10^{69}$ |
| $^{192}Os \to {}^{188}W$ | 361(4) | 40.86(5) | $0^+$, g.s. **) | $\geq 5.8 \times 10^{18}$ | $1.4 \times 10^{140} - 1.7 \times 10^{153}$ |
| | | | $2^+$, 143.2 | $\geq 2.7 \times 10^{19}$ | $9.9 \times 10^{190} - 1.6 \times 10^{215}$ |

*) A γ-peak with energy 125.4 keV is expected in β decay of $^{185}W$, the progeny of $^{189}Os$.

**) A γ-peak with energy 155.0 keV is expected in β decay of $^{188}Re$, the progeny of β active $^{188}W$ produced in the α decay of $^{192}Os$.

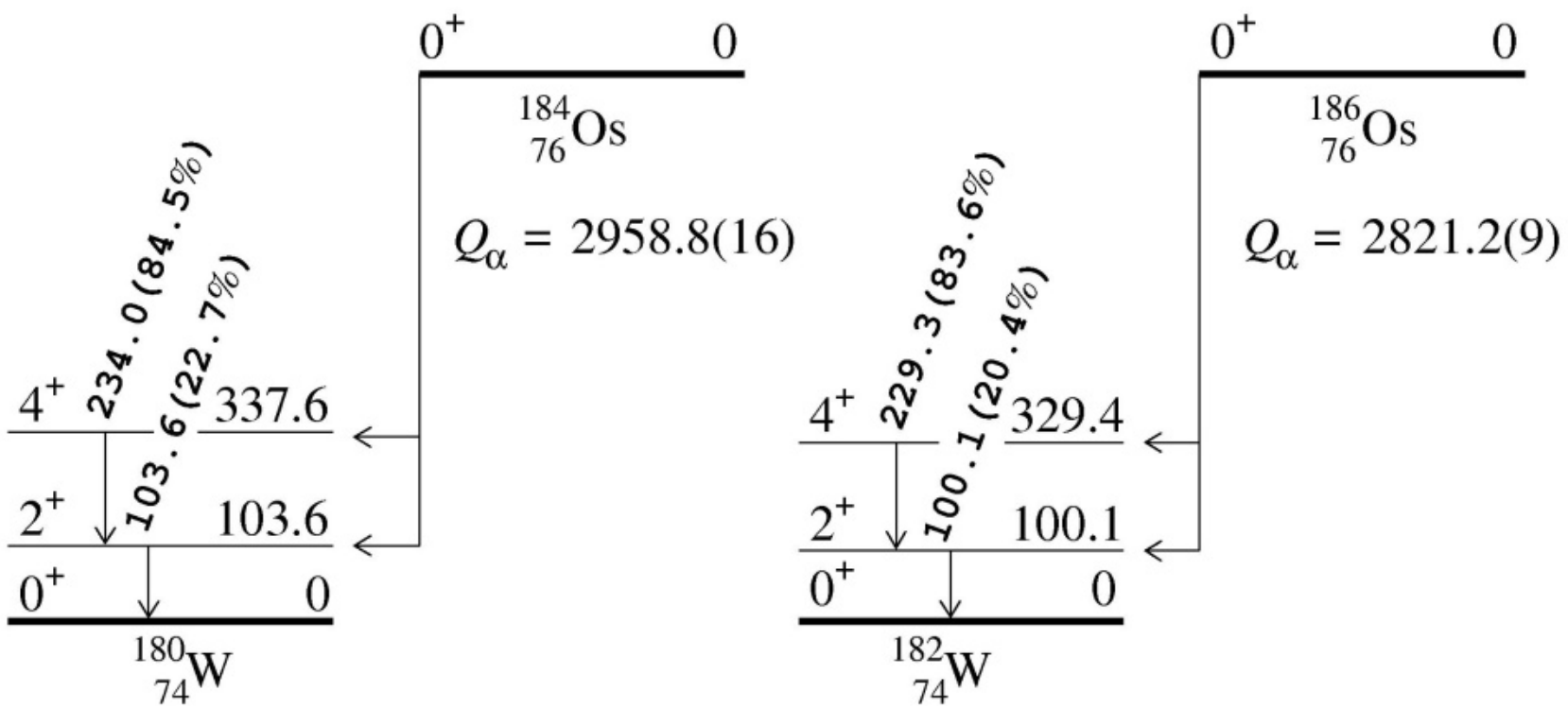

Fig. 1. Simplified schemes of $^{184}$Os and $^{186}$Os α decay to the two lowest excited levels of their progeny. The $Q_\alpha$ values, energy of the levels and of the de-excitation γ quanta are given in keV. The probabilities of γ quanta emission, assuming 100% populations of the corresponding level, taking into account the total internal conversion coefficients [17, 18], are given in parentheses.

Double beta (2β) decay is an extremely important and interesting scope of research in nuclear and astroparticle physics both from the experimental and theoretical points of view. The neutrinoless mode of the 2β decay (0ν2β) is a promising tool to check the lepton number conservation, to clarify the nature of the neutrino (Dirac or Majorana particle), to determine an absolute scale of the neutrino mass (assuming the light neutrino exchange model of the decay) and the neutrino mass hierarchy (utilizing also a global analysis of different neutrino experiments), and to test many other effects beyond the Standard Model (SM) of particles and interactions [19, 20, 21, 22, 23]. While the two-neutrino mode of the decay (2ν2β$^-$) is allowed in the SM and was detected for 11 nuclides with the half-lives in the range of $T_{1/2} \sim 10^{18} - 10^{24}$ yr [24, 25], the 0ν2β decay is forbidden in the SM since it violates the lepton number conservation and is still unobserved. The currently the best sensitivity of the existing 0ν2β$^-$ experiments is at the level of lim $T_{1/2} \sim 10^{24} - 10^{26}$ yr [23], whereas the sensitivity to the 2β processes increasing the atomic number ("double-beta plus" decay), such as the double electron capture (2EC), the electron capture with positron emission (ECβ$^+$), and the double positron (2β$^+$) decay, is substantially lower both for the 2ν and 0ν modes, at the level of lim $T_{1/2} \sim 10^{20} - 10^{22}$ yr [26].

Osmium contains two potentially 2β active isotopes: $^{184}$Os with the energy release $\beta_{EC}$ = 1452.9(7) keV and $^{192}$Os with $Q_{2\beta}$ = 406(3) keV [15]. The 2β decay schemes of $^{184}$Os and $^{192}$Os are presented in Fig. 2 and 3, respectively. A near resonant enhancement is expected for neutrinoless double electron capture in $^{184}$Os thanks to the mass degeneracy of the ground state and of the $(0)^+$ 1322.2-keV and $2^+$ 1431.0-keV excited levels of $^{184}W$. Assuming the effective Majorana neutrino mass $\langle m_\nu \rangle = 1$ eV, the half-life of the 0νKL transition to the $(0)^+$ 1322.2 keV excited level of $^{184}W$ was estimated twice as $T_{1/2} = (0.7 - 2) \times 10^{27}$ yr [27] and $T_{1/2} > 1.3 \times 10^{29}$ yr [28]. In [26] the half-life of $^{184}Os$ relative to the 0ν2L transition to the $2^+$ 1431.0-keV excited level of $^{184}Os$ was calculated as $T_{1/2} = 4.5 \times 10^{30} - 4.5 \times 10^{35}$ yr for $\langle m_\nu \rangle = 1$ eV. For $^{192}Os$ theoretical predictions were reported only for the 2β decays to the ground state of $^{192}Pt$ in the range of $T_{1/2} = 1.3 \times 10^{24} - 4.6 \times 10^{29}$ yr for the 2ν mode [29, 30, 31, 32, 33, 34] and $T_{1/2} = (4.1 \times 10^{24} - 3.3 \times 10^{26})$ yr [29, 35] for the neutrinoless decay (assuming $\langle m_\nu \rangle = 1$ eV). Certainly, the transition to the $2^+$ 316.5 keV excited level is expected to be suppressed additionally by the smaller phase space and the spin change by two units.

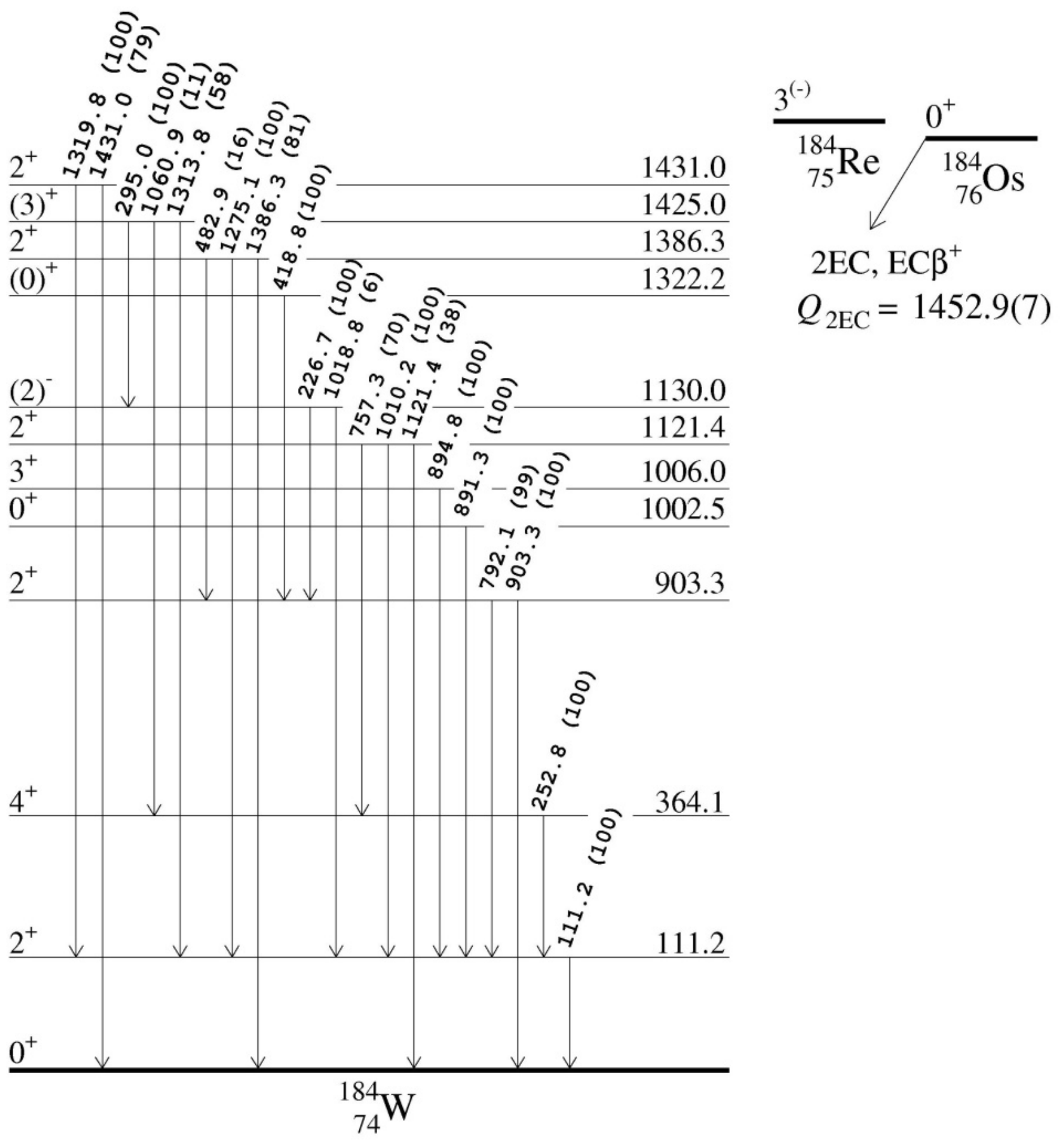

Fig. 2. The simplified 2β decay scheme of $^{184}Os$ [36]. The value of $Q_{2EC}$ for $^{184}Os$ is from [15]. The energies of the excited levels, of the emitted γ quanta and the decay energy $Q_{2EC}$ are given in keV. The relative intensities of the γ-ray transitions (given in parentheses) are relative branching ratios from each level.

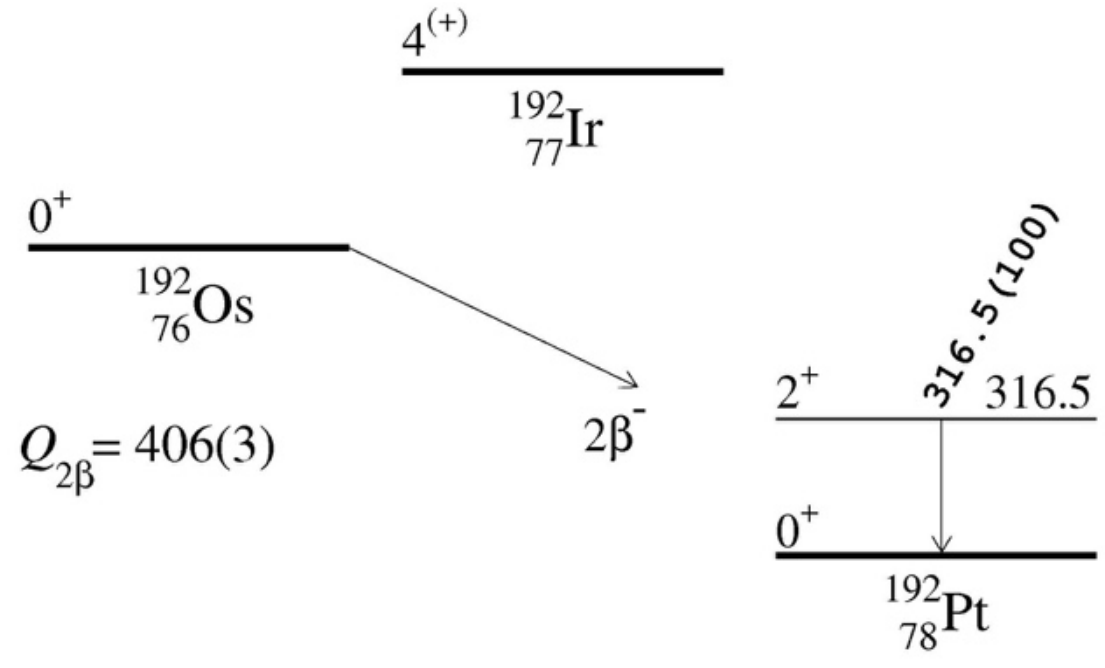

Fig. 3. The decay scheme of $^{192}$Os [37]. The $Q_{2\beta}$ value is from [15]. The energy of the excited level, of the emitted γ quanta and the decay energy $Q_{2\beta}$ are in keV.

A possibility of nuclear decay with simultaneous emission of two alpha particles (2α decay), (A, Z) → (A–8, Z–4) + 2α, was theoretically considered in the 1980's [38, 39, 40, 41]. In these studies, the expected half-lives were estimated for several nuclides. There were no further theoretical and experimental activities in this field during the following forty years. Recently, it was re-considered in [42], with theoretical calculations of the expected half-lives for naturally occurring radionuclides, and the first experimental half-life limit was set for $^{209}$Bi as $T_{1/2} > 2.9 \times 10^{20}$ yr. In approaches where the two emitted α particles are considered as a $^{8}$Be cluster [39, 40, 41, 42, 43, 44, 45], the calculated $T_{1/2}$ values are very long, e.g. in [42] for the stable or long-lived radionuclides, they are at level of $10^{33}$ yr or higher. However, in the Refs. [46, 47, 48] the so-called symmetric 2α decay was considered, when two α particles are emitted in opposite directions with equal energies $Q_{2\alpha}/2$. Microscopic calculations [46] of $T_{1/2}$ for such a process gave values much lower than those in [42] (where the semi-empirical formulas of Ref. [49] for cluster decay were used): $7.3 \times 10^{10}$ yr instead of $2.1 \times 10^{31}$ yr for $^{212}$Po, and $5.5 \times 10^{6}$ yr instead of $2.3 \times 10^{20}$ yr for $^{224}$Ra. In Ref. [47] even a much shorter half-life of only 2.6 yr was obtained for $^{224}$Ra in a phenomenological treatment of the symmetric 2α decay. Thus, from the theoretical side, the situation in $2\alpha$ decay studies is rather intriguing. Experimental investigations of this process could help to clarify the picture.

All seven naturally occurring osmium isotopes are theoretically unstable with respect to $2\alpha$ decay with energy releases ranging from 767 keV to 5474 keV. The isotopes are listed in Table 2 together with the energy release and the theoretical predictions obtained in different approaches.

One can see that the theoretical estimations give half-life values which differ by many orders of magnitude.

Table 2. Possible 2α decays of naturally occurring osmium isotopes. $^{189}$Os and $^{192}$Os have further emissions of γ quanta; given are the energy release ($Q_{2\alpha}$) [15] and the theoretical calculations of $T_{1/2}$ in different approaches.

| Nuclide | $Q_{2\alpha}$ (keV) | Theoretical $T_{1/2}$ (yr) | | | |
|---|---|---|---|---|---|
| | | [42]* | [43] | [44] | [47]** |
| $^{184}$Os→$^{176}$Hf | 5474.0(17) | $1.4 \times 10^{57}$ | $9.8\times10^{53}$–$1.5\times10^{61}$ | $2.1\times10^{50}$–$9.1\times10^{57}$ | $7.3\times10^{19}$ |
| $^{186}$Os→$^{178}$Hf | 4585.5(16) | $2.9 \times 10^{70}$ | $2.3\times10^{72}$–$1.8\times10^{80}$ | $1.9\times10^{63}$–$3.8\times10^{77}$ | $7.8\times10^{26}$ |
| $^{187}$Os→$^{179}$Hf | 4394.2(16) | $6.9 \times 10^{73}$ | $2.2\times10^{70}$–$1.6\times10^{78}$ | – | $4.5\times10^{28}$ |
| $^{188}$Os→$^{180}$Hf | 3792.4(16) | $1.5 \times 10^{86}$ | $3.2\times10^{82}$–$5.5\times10^{90}$ | $4.3\times10^{78}$–$4.2\times10^{88}$ | $1.6\times10^{35}$ |
| $^{189}$Os→$^{181}$Hf | 3566.3(15) | $4.0 \times 10^{91}$ | $7.1\times10^{87}$–$1.7\times10^{96}$ | – | $1.1\times10^{38}$ |
| $^{190}$Os→$^{182}$Hf | 2492(6) | $1.1 \times 10^{127}$ | $8.7\times10^{122}$–$1.0\times10^{132}$ | $8.6\times10^{117}$–$1.8\times10^{130}$ | $5.8\times10^{56}$ |
| $^{192}$Os→$^{184}$Hf | 766(40) | $2.1 \times 10^{301}$ | $1.9\times10^{276}$–$3.8\times10^{282}$ | $1.2\times10^{278}$–$1.7\times10^{281}$ | $5.0\times10^{148}$ |

* For $^{188,190,192}$Os, calculated here following [42].

** Calculated here following Ref. [47] supposing a spherical shape of the nuclei.

It should be noted that the half-life of $^{184}$Os, calculated with the approach of Ref. [47] in the hypothesis of a spherical shape of the nucleus, is "only" $T_{1/2} = 7.3 \times 10^{19}$ yr. Thus, it looks realistic to investigate this decay using currently available experimental techniques: the half-life is close to the $T_{1/2}$ values of nuclides for which single α decays were discovered recently: $1.9 \times 10^{19}$ yr for $^{209}$Bi [50], $1.1 \times 10^{18}$ yr for $^{180}$W [51], $5.0 \times 10^{18}$ yr for $^{151}$Eu [52], and $1.7 \times 10^{21}$ yr for the $^{209}$Bi decay to the first excited level of $^{205}$Tl [53]. Certainly, one should apply an experimental technique able to register α particles. Unfortunately, the rather low natural abundance of $^{184}$Os (~0.02%, see Table 1) will be a challenge for the observation of this decay. However, osmium isotopes, particularly $^{184}$Os, can be enriched by centrifugation (see discussion in [54]). Thus, one should keep in mind this interesting nuclide.

In our measurements with γ-ray spectrometry, where Os is an external sample, we can look for the 2α decays of $^{189}$Os and $^{192}$Os. In the 2α decay $^{189}$Os → $^{181}$Hf the progeny is β unstable and decays to $^{181}$Ta with $T_{1/2}$ = 42.39 d [55] and $Q_\beta$ = 1036.1 keV [15]. The energy of the most intense γ quanta emitted in the β decay of $^{181}$Hf is 482.2 keV with a yield of 80.5%. A simplified decay scheme of the 2α decay of $^{189}$Os with further β decay of $^{181}$Hf is shown in Fig. 4.

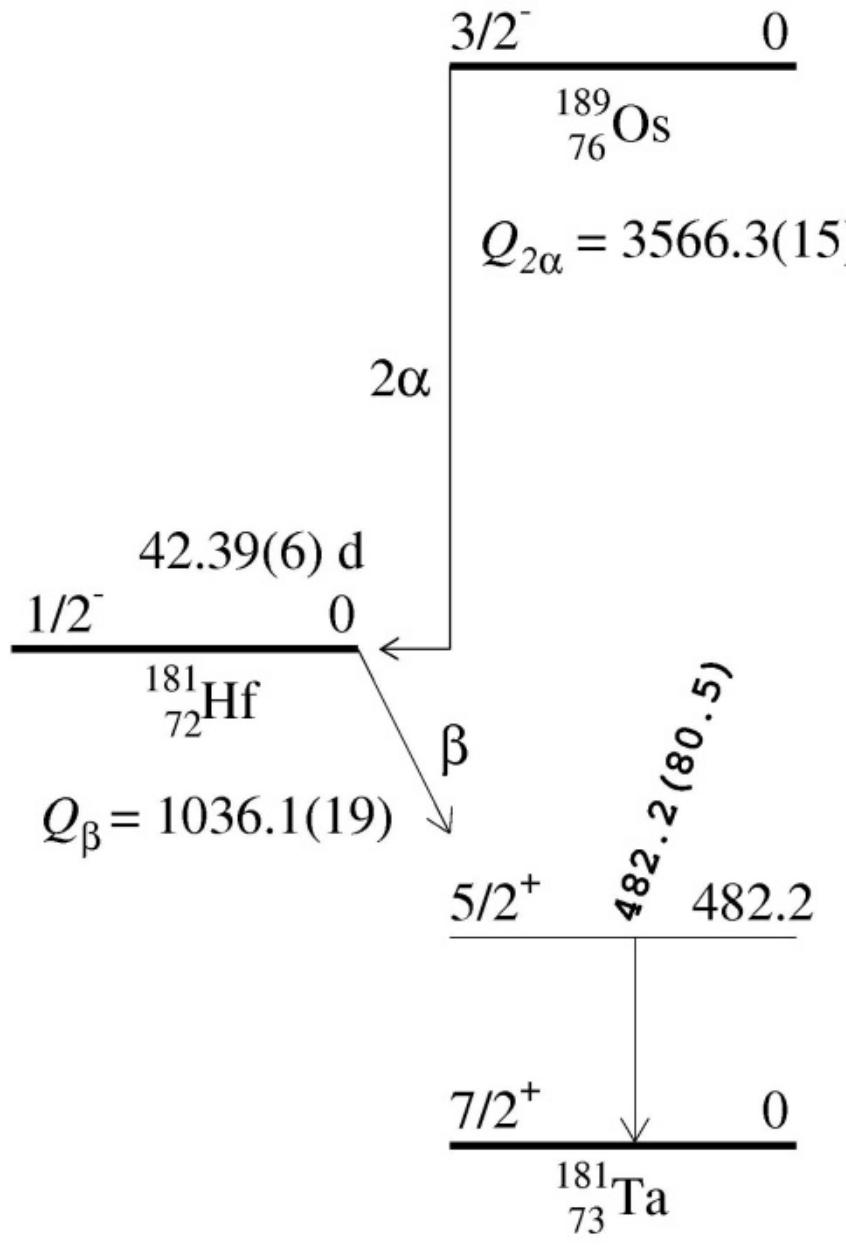

Fig. 4. A simplified 2α decay scheme of $^{189}$Os [55]. The $Q_\beta$ value for $^{181}$Hf is from [15]. The energies of the excited 5/2$^+$ level of $^{181}$Ta and of the de-excitation γ quanta are given in keV. The absolute intensity of the γ-ray transition is given in parentheses.

For the transition $^{192}$Os → $^{184}$Hf the progeny is unstable and two further $\beta^-$ decays should follow [36]: $^{184}$Hf ($T_{1/2}$ = 4.12 h, $Q_\beta$ = 1340 keV) → $^{184}$Ta ($T_{1/2}$ = 8.7 h, $Q_\beta$ = 2866 keV) → $^{184}$W (stable). A simplified decay scheme of $^{192}$Os is shown in Fig. 5.

In the 2α decay of $^{190}$Os the progeny $^{182}$Hf has a too long half-life, $T_{1/2} \approx 9\times10^6$ yr, which makes impractical its investigation.

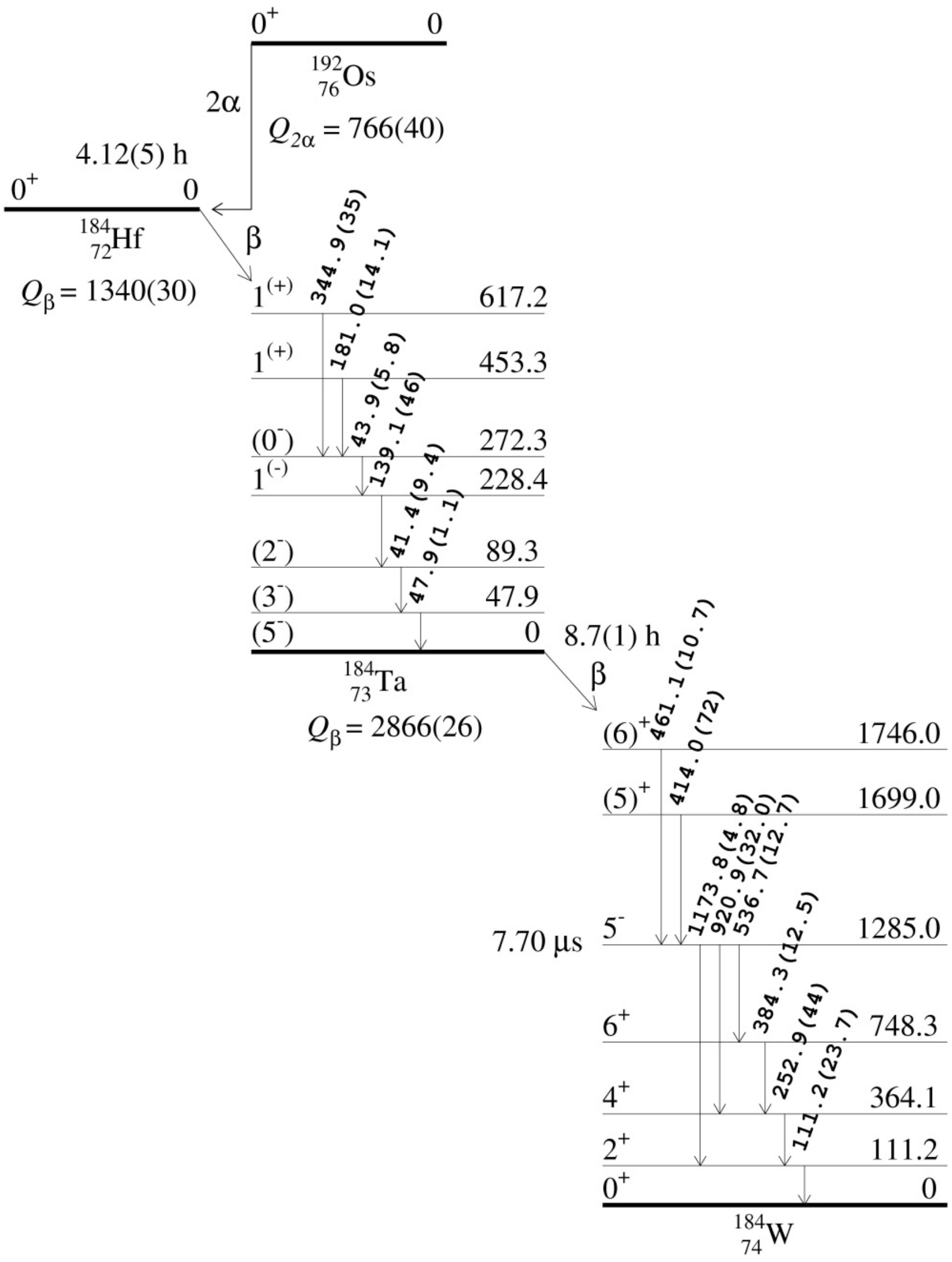

Fig. 5. The simplified 2α decay scheme of $^{192}$Os [36]. The energies of the excited levels of $^{184}$Ta and $^{184}$W, and of the de-excitation γ quanta are given in keV. The absolute intensities of the γ-ray transitions are given in parentheses.

## 2. Measurement

### 2.1. The first stage: γ-ray spectrometry of an osmium sample in form of ingots

Four ingots of ultrapure osmium of 99.999% purity with a total mass of 172.50 g (see Fig. 6) were used in the first stage of the measurement. The material was obtained using electron-beam melting of osmium powder with further purification by electron-beam zone refining at the National Science Center “Kharkiv Institute of Physics and Technology” of the National Academy of Sciences of Ukraine.

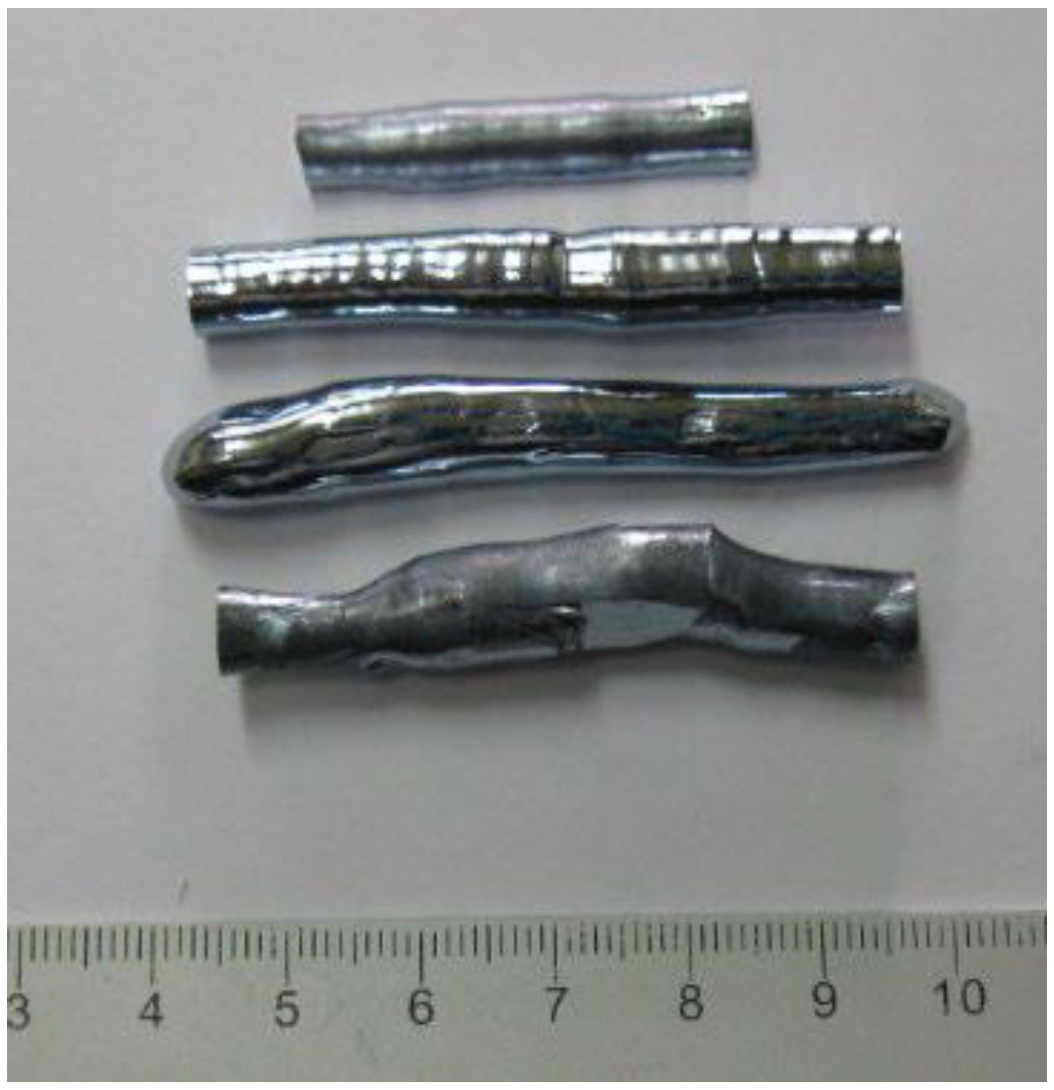

Fig. 6. A photograph of the high-purity osmium ingots with a total mass 172.50 g utilized at the first stage of the measurement.

Low-background measurements with the osmium sample were performed over 2741 h at the SubTerranean Low Level Assay (STELLA) laboratory of the Gran Sasso National Laboratory of the INFN (Italy), at a depth of ~3.6 km of water equivalent, using the ultra-low background semi-coaxial high purity germanium (HPGe) detector *GeCris* with a volume of 465 $cm^3$. The detector was shielded by low radioactive copper (≈ 5 cm) and lead (≈ 25 cm). The set-up was flushed with high purity boil-off nitrogen to remove environmental radon. Some characteristics of the set-up are presented in Table 3.

Table 3. Characteristics of the low-background set-ups utilized in the experiment.

| Stage of the experiment | 1 | 2 | 3 |
|---|---|---|---|
| Detector, type, total volume | HPGe *GeCris*, p-type, 465 $cm^3$ | BEGe, p-type, 118 $cm^3$ | HPGe *GS1*, p-type, 275 $cm^3$ |
| Window material, thickness | Cu, 1.0 mm | Al, 1.5 mm | Cu, 1.5 mm |
| Dead layer on top of the Ge crystal (used in the Monte Carlo simulations of the set-ups) | 1.0 mm | 0 mm | 1.74 mm |
| Mass of the osmium sample (g) | 172.50 | 117.96 | 58.78 |
| Number of $^{184}$Os ($N^{184}$) and $^{186}$Os ($N^{186}$) nuclei in the sample | $N^{184} = 9.29 \times 10^{19}$<br>$N^{186} = 8.6872 \times 10^{21}$ | $N^{184} = 6.35 \times 10^{19}$<br>$N^{186} =5.9405 \times 10^{21}$ | $N^{184} = 3.16 \times 10^{19}$<br>$N^{186} =2.9602 \times 10^{21}$ |
| Time of measurements (h) | 2741 | 15851 | 23840 |
| Background counting rate in the energy interval 95−110 keV ($d^{-1}$ $keV^{-1}$) | 0.606(19) | 0.605(8) | 1.125(9) |
| Energy resolution for γ quanta at 100 keV (FWHM, keV) | 1.02(8) | 0.86(5) | 1.17(3) |
| Full absorption peak detection efficiency for γ quanta 103.6 keV ($\varepsilon^{103.6}$) and 100.1 keV ($\varepsilon^{100.1}$) simulated by Monte Carlo method | $\varepsilon^{103.6}$ = 0.199%<br>$\varepsilon^{100.1}$= 0.174% | $\varepsilon^{103.6}$ = 1.382%<br>$\varepsilon^{100.1}$ = 1.274% | $\varepsilon^{103.6}$ = 1.360%<br>$\varepsilon^{100.1}$ = 1.187% |

The energy spectrum acquired over 2741 h with the osmium ingots is shown in Fig. 7. The dependence of the detector energy resolution ($R^1_\gamma$, full width at half of maximum, FWHM, keV) on γ quanta energy ($E_\gamma$, keV) in the energy spectrum taken over 2741 h was determined using the γ peaks of $^{210}$Pb (46.5 keV), $^{212}$Pb (238.6 keV), $^{214}$Pb (351.9 keV) and $^{214}$Bi (609.3 keV) as:

$$R_\gamma^1 = \sqrt{0.76(15) + 0.0028(5) \times E_\gamma} \ . \quad (1)$$

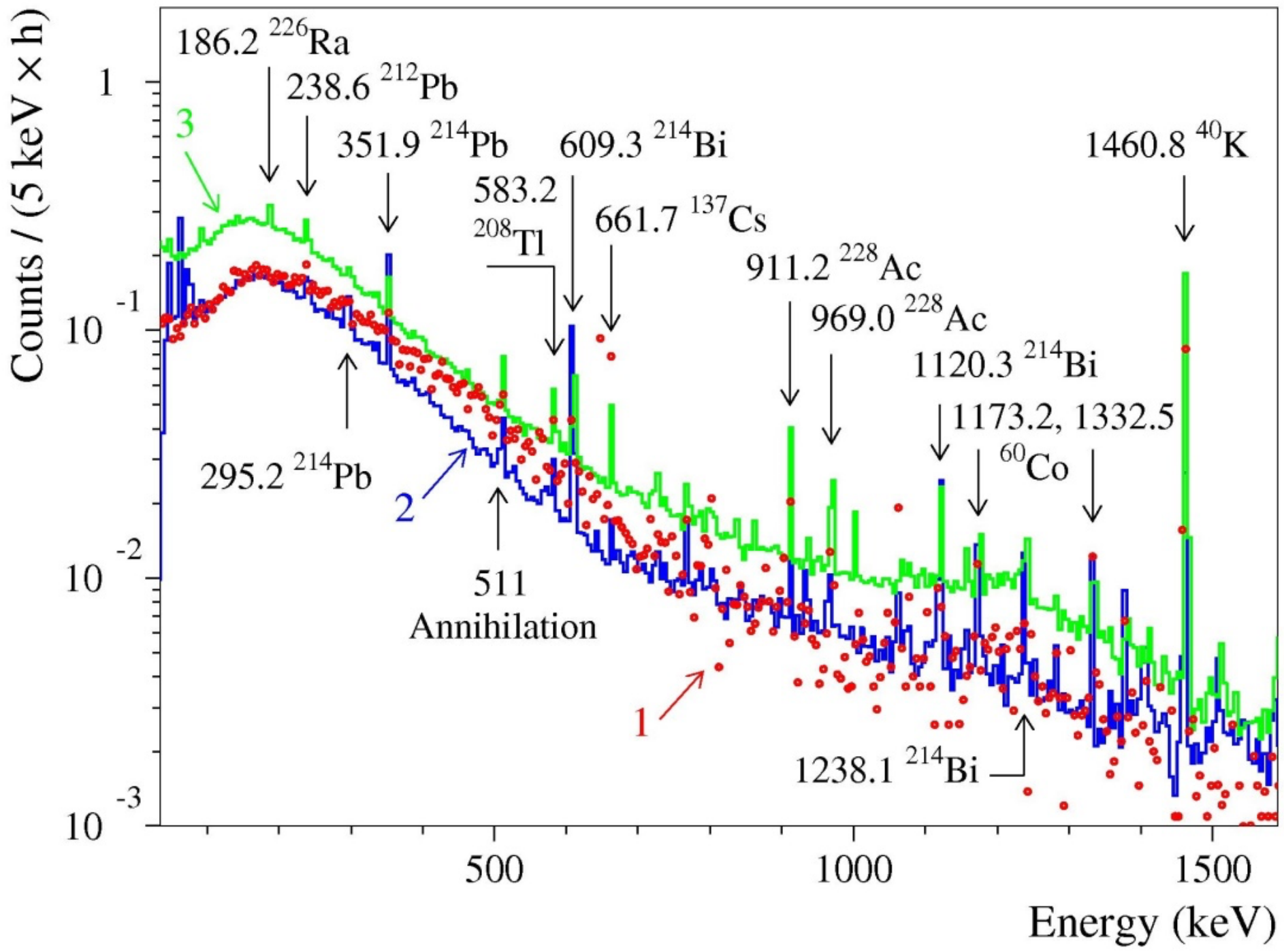

Fig. 7. Energy spectra measured in the 1st (over 2741 h, red dots, labelled as “1”), in the 2nd (15851 h, blue histogram, “2”) and in the 3rd (23840 h, green histogram, “3”) stages of the measurement with osmium samples. The spectra are normalized to the time of measurements. Energies of γ peaks are given in keV.

All the peaks in the energy spectrum measured with the osmium were attributed to γ quanta of $^{40}$K, $^{60}$Co, $^{137}$Cs, $^{207}$Bi, $^{232}$Th and $^{238}$U with their relative progeny. The $^{137}$Cs pollution of the material can be explained by sequences of the Chernobyl accident. Cosmogenic $^{185}$Os (decays by electron capture with the decay energy $Q_{EC}$ = 1013.1 keV and the half-life $T_{1/2}$ = 93.6 d) was observed too. The high radiopurity level of the osmium material estimated by analyzing the γ peaks in the data is presented in Table 4. More detail of the 1st stage of the measurement and results of the first search for 2β decay processes in $^{184}$Os and $^{192}$Os have already been reported in [56].

Table 4. Radioactivity contaminations of the osmium sample assayed by low-background γ-ray spectrometry (values given in mBq $kg^{-1}$) for the $1^{st}$ and $2^{nd}$ stages of the measurement. The upper limits are presented at 90% C.L., the uncertainties are given with 68% C.L. The activity of $^{207}$Bi (reported in [56] as $0.4 \pm 0.1$ mBq $kg^{-1}$) was replaced in the present study by limit taking into consideration the intrinsic background of the GeCris detector.

| Chain | Nuclide | Stage of the measurement, Reference | |
|---|---|---|---|
| | | 1 [56] | 2 [54] |
| | $^{40}$K | ≤ 1.9 | 11 ± 4 |
| | $^{60}$Co | ≤ 0.1 | ≤ 1.3 |
| | $^{137}$Cs | 1.9 ± 0.3 | 0.5 ± 0.1 |
| | $^{185}$Os | 3.0 ± 0.3 | – |
| | $^{207}$Bi | ≤ 0.6 | – |
| | $^{241}$Am | – | ≤ 5.6 |
| $^{232}$Th | $^{228}$Ra | ≤ 2.0 | ≤ 6.6 |
| | $^{228}$Th | ≤ 2.3 | ≤ 16 |
| $^{235}$U | $^{235}$U | | ≤ 8.0 |
| | $^{231}$Pa | | ≤ 3.5 |
| | $^{227}$Ac | | ≤ 1.1 |
| $^{238}$U | $^{238}$U | | ≤ 35 |
| | $^{226}$Ra | ≤ 0.6 | ≤ 4.4 |
| | $^{210}$Pb | | ≤ 180 |

## 2.2. The second stage: measurements of osmium in the form of thin plates

Taking into account a strong self-absorption of the low energy γ quanta expected in the α decays of $^{184}$Os and $^{186}$Os to the first excited levels of their progeny in the osmium sample, the osmium ingots were cut into 0.79 – 1.25 mm thick plates by electroerosion cutting with a brass wire in kerosene.

The low-background measurements with the thin osmium sample were carried out with an ultra-low background broad-energy germanium (BEGe) detector (with a total volume of 118 cm$^3$) that has the distinct advantage of enhanced detection efficiency for low energy γ quanta. The thin osmium plates with a mass of 117.96 g were placed on the top and around the detector endcap made in aluminum alloy. The detector was shielded with layers of high-purity copper (≈5 cm) and 20 cm of lead. The energy spectrum measured with the sample over 15851 h is shown in Fig. 7. Energy resolution in the 2$^{nd}$ stage can be approximated by the function [16]:

$$R_\gamma^2 = 0.57(5) + 0.029(2) \times \sqrt{E_\gamma}. \qquad (2)$$

The main characteristic of the set-up in the 2$^{nd}$ stage are given in Table 3.

The radioactivity concentration of the osmium material measured in this 2$^{nd}$ stage is reported in Table 4. The slight increase of the $^{40}$K activity can be explained by the cutting procedure, while the decrease of $^{137}$Cs activity is achieved thanks to a more accurate cleaning of the material. Cosmogenic $^{185}$Os was not present in this 2$^{nd}$ stage because of the rather long cooling time underground (about 3 years) before the 2$^{nd}$ stage measurements started.

Considering a very big uncertainty of the $^{184}$Os recommended isotopic abundance, $\delta$ = 0.02(2)% [57], the isotopic composition of the osmium was analyzed at the John de Laeter Centre of the Curtin University (Perth, Western Australia) with high precision using negative thermal ionization mass spectrometry [16]. The accuracy of the measured isotopic concentrations of the sample is much higher than that of the adopted reference values, especially for $^{184}$Os. The measured isotopic concentrations of the osmium are given in Table 1.

The results of the 2$^{nd}$ stage of the measurement to search for α decays of the naturally occurring osmium nuclides have already been reported in [16], while the results of the search for the 2β decay processes in $^{184}$Os and $^{192}$Os can be found in [54].

### 2.3. The third stage: thin osmium plates placed directly on the Ge crystal

In the last stage of the measurement, aiming at improving further the detection efficiency to low energy γ quanta, part of the osmium plates with the total mass of 58.78 g were glued to a thin plastic plate (∅70.5 × 0.8 mm) using Armstrong epoxy A-12 and were mounted directly on the Ge crystal of the HPGe semi-coaxial p-type detector GS1 with a total volume of 275 $cm^3$ similar to what has been described in [58] for a platinum sample. A photograph of the sample and a simplified scheme of the set-up are presented in Fig. 8. The detector was shielded with high purity copper (5 cm) and a graded lead shield (15 cm); the set-up was flushed with high purity boil-off nitrogen to remove environmental radon. The data with the osmium sample were acquired over 23840 h, while the background of the detector without sample was taken for 5292 h. The energy spectrum acquired in the $3^{rd}$ stage of the measurement with the osmium sample is shown in Fig. 9 together with the background data. The spectrum with the sample is almost indistinguishable from the background data that confirms the high radiopurity level of the osmium sample. However, as one can see in Fig. 7 (see also in Table 3), the background counting rate of the detector in the $3^{rd}$ stage is slightly higher than those in the two previous stages.

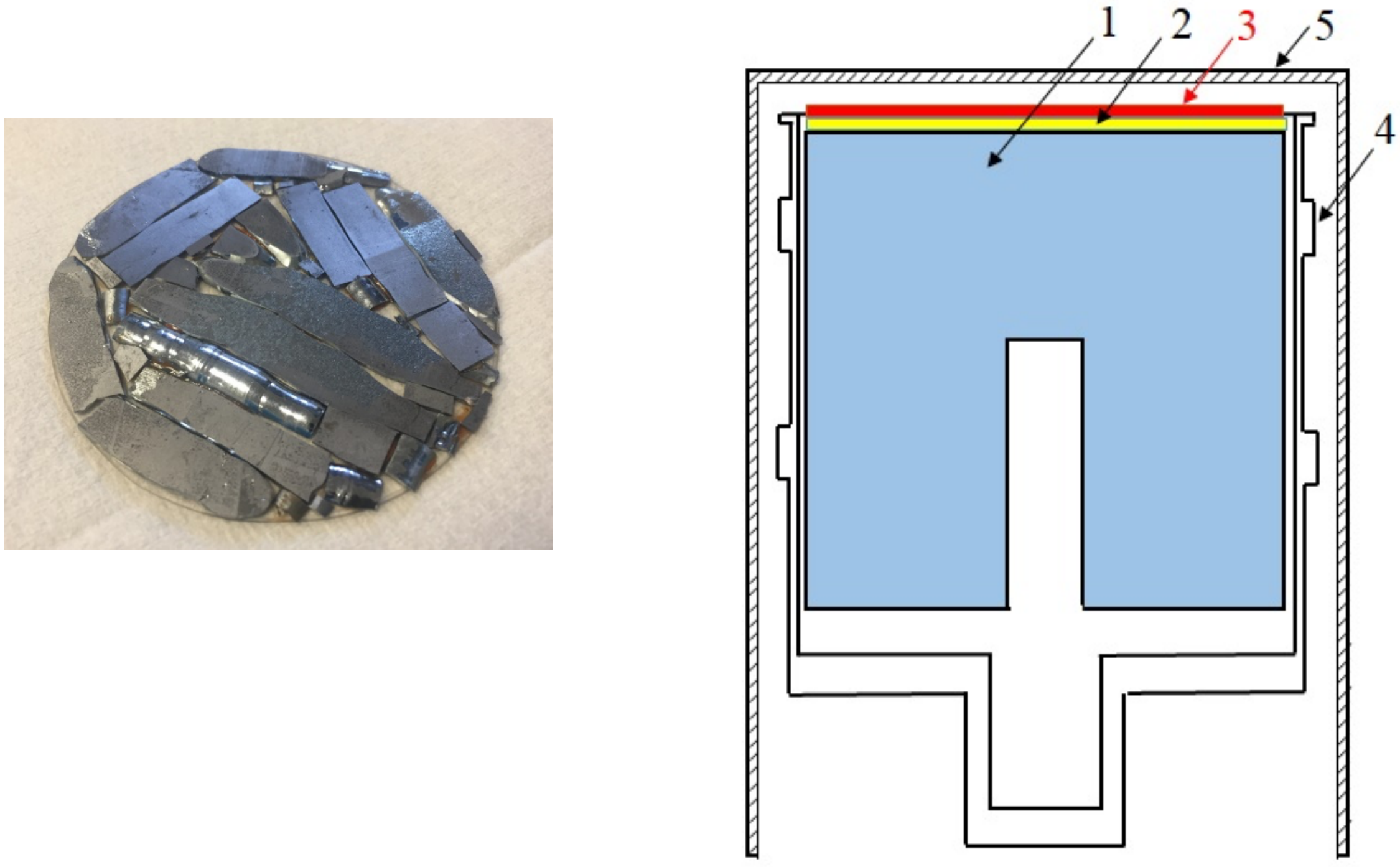

Fig. 8. A photograph of the osmium plates glued to a thin plastic support (left panel) and a simplified scheme of the set-up (right panel): (1) Ge crystal, (2) plastic support, (3) osmium metal plates, (4) Ge crystal holder, (5) detector endcap.

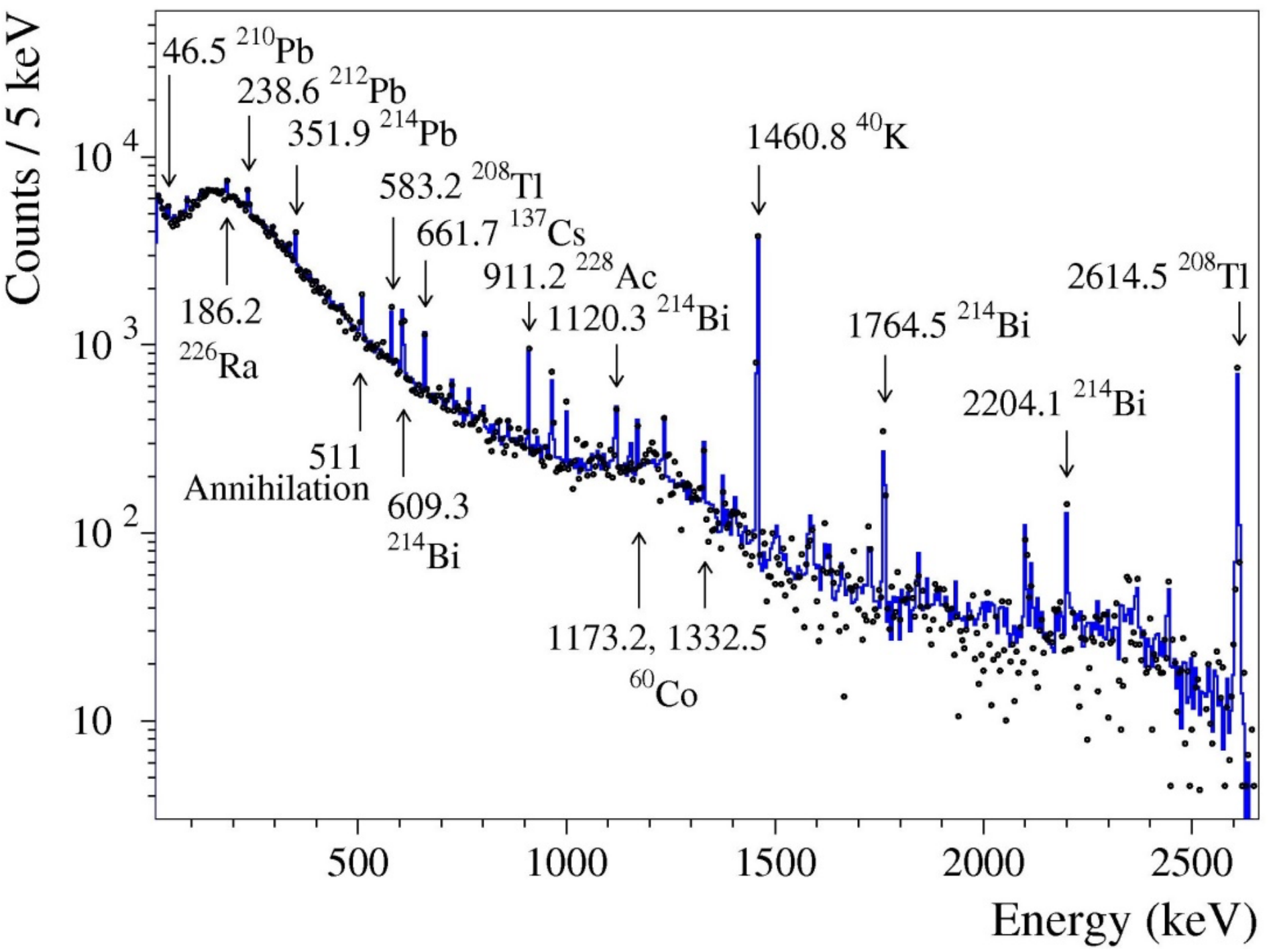

Fig. 9. Energy spectrum measured in the 3rd stage of the measurement with the HPGe-detector GS1 for 23840 h with osmium sample (blue histogram) and background spectrum without sample taken over 5292 h (black dots). The background data are normalized on the time of measurements with the sample.

The energy resolution of the detector (FWHM) in the low energy part of the final spectrum was estimated using background γ peaks from $^{232}$Th and $^{238}$U chains, and $^{137}$Cs: 46.5 keV ($^{210}$Pb), 143.8 keV ($^{235}$U), 238.6 keV ($^{212}$Pb), 338.3 keV ($^{228}$Ac), 351.9 keV ($^{214}$Pb), 583.2 keV ($^{208}$Tl), 609.3 keV ($^{214}$Bi), 661.7 keV ($^{137}$Cs) as

$$R_{\gamma}^{3} = \sqrt{1.15(6) + 0.00222(15) \times E_{\gamma}} \, . \tag{3}$$

The characteristics of the low-background set-ups used in the three stages of the measurement are summarized in Table 3. The full absorption peak detection efficiencies to γ quanta were calculated using the EGSnrc simulation package [59] and the event generator DECAY0 [60]. Despite it would seem better to place the sample inside the cryostat of the detector, one can see that this measure did not significantly improve the sensitivity of the measurement to low-energy γ quanta due to the rather thick dead layer of the *GS1* detector in comparison to that of the BEGe.

## 3. Results and discussion

### 3.1. Search for α decay of the osmium nuclides with emission of γ quanta

The α transitions of $^{184}Os$ and $^{186}Os$ to the first excited levels of their progeny were not observed in the data of the 1st stage of the measurement. However, taking into account the presence of a peculiarity at 103(1) keV with an area $S = (22 \pm 14)$ counts, an indication of the α decay of $^{184}Os$ to the $2^+$ 103.6-keV excited level of $^{180}W$ with a half-life $T_{1/2}(^{184}\mathrm{Os} \rightarrow 103.6) = 5^{+8}_{-2} \times 10^{14}$ yr was reported. For the α decay of $^{186}Os$ to the $2^+$ 100.1-keV excited level of $^{182}W$ a half-life limit $T_{1/2}(^{186}\mathrm{Os} \rightarrow 100.1) \geq 4 \times 10^{16}$ yr was set [61]. However, the indication of the α decay of $^{184}Os$ to the first excited level of $^{180}W$ was not confirmed in the further, more sensitive stages of the measurement. Thus, we reanalyze the data assuming no effect in the spectrum taken for 2741 h in the 1st stage of the measurement with the osmium ingots. Half-life limits for α transitions with emission of γ quanta can be calculated with the following formula:

$$\lim T_{1/2} = \frac{N \cdot \ln 2 \cdot \eta \cdot \varepsilon \cdot t}{\lim S}, \qquad (4)$$

where $N$ is the number of nuclei of the isotope of interest, $\eta$ is the γ quanta emission intensity, $\varepsilon$ is the full absorption peak detection efficiency for γ quanta expected in the decay, $t$ is the time of measurement, and lim $S$ is a lower limit on the number of events of the effect searched for which can be excluded at a given confidence level (C.L.)[1].

A part of the energy spectrum measured in the 1st stage of the measurement in the vicinity of the energies expected in the α decays of $^{184}Os$ and $^{186}Os$ to the first excited levels of their progeny is presented in Fig. 10. The results of fits in the energy region 93.6–112.7 keV by a simple background models built from a linear function to describe continuous distribution and Gaussians at 100.1 keV or at 103.6 keV with the width fixed according to Eq. (1) are presented in Fig. 10 too. The quality of the fits is good: $\chi^2$/n.d.f. = 41.9/51 = 0.822 for the 103.6-keV peak and $\chi^2$/n.d.f. = 41.5/51 = 0.814 for 100.1-keV peak. The fits returned the peaks areas $S^{103.6}$ = (−0.8 ± 8.2) counts and $S^{100.1}$ = (3.3 ± 8.4) counts, i.e. there is no evidence for an effect. According to the recommendations in [62] the excluded peaks areas are lim $S^{103.6}$ = 12.7 counts and lim $S^{100.1}$ = 17.1 counts. The excluded peaks are shown in Fig. 10. Taking into account the Monte Carlo simulated detection efficiencies for the γ quanta with the expected energies (see Table 3), the relative emission probabilities for the γ quanta of the corresponding level (see Fig. 1) and the numbers of nuclei in the osmium sample (measured by the mass spectrometry and given in Table 3) one obtains the following half-life limits: $T_{1/2}$ ($^{184}$Os → 103.6) ≥ 7.2 × $10^{14}$ yr and $T_{1/2}$ ($^{186}$Os → 100.1) ≥ 3.9 × $10^{16}$ yr.

[1] All the area and half-life limits in the present study are given at 90% C.L.

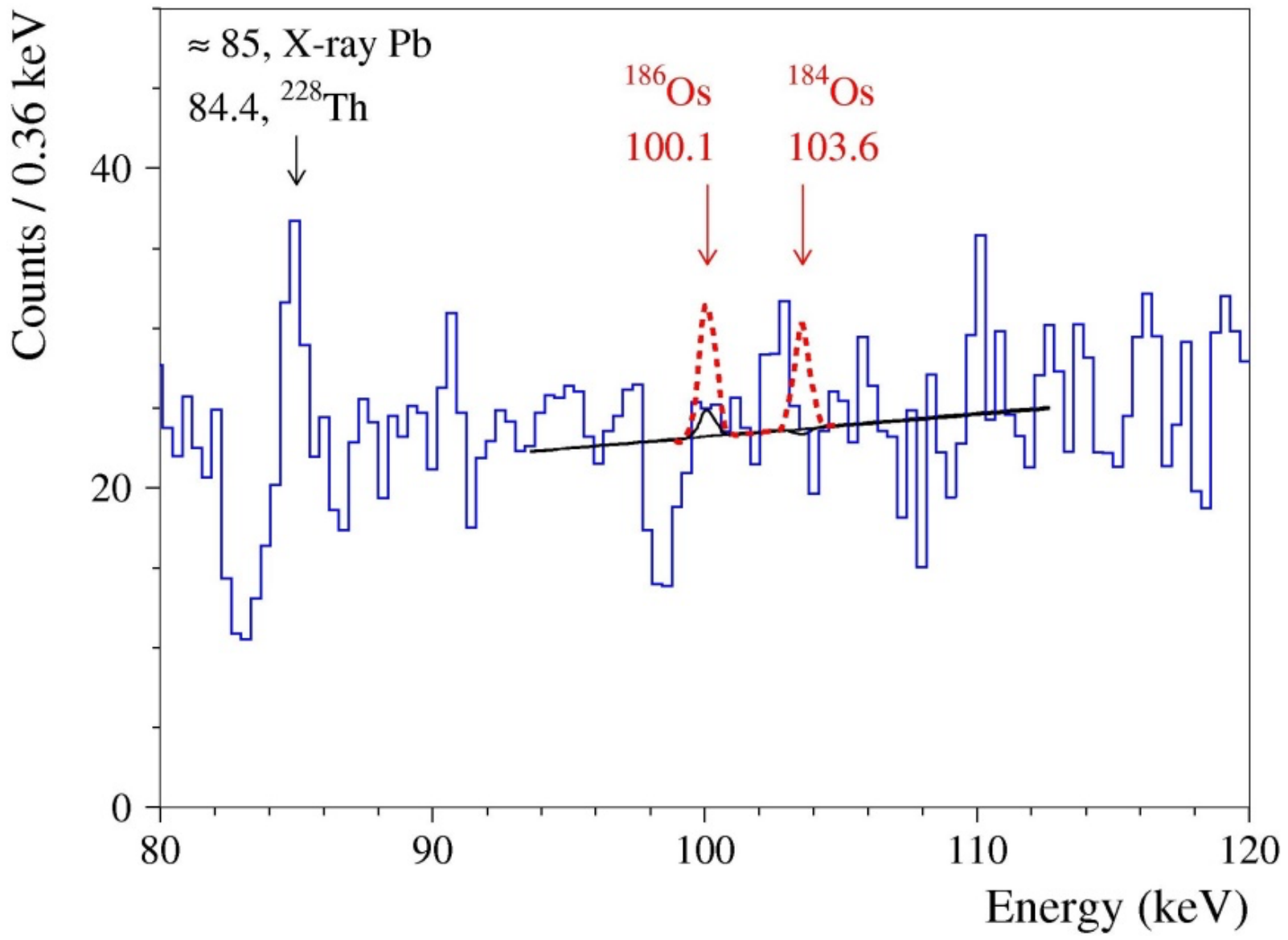

Fig. 10. Part of the energy spectrum measured for 2741 h in the first stage of the measurement with the osmium ingots in the region where the 100.1-keV and 103.6-keV peaks after the α transitions of $^{186}$Os and $^{184}$Os to the first excited levels of the progeny nuclei are expected. The fits of the data with the background model (see text) are shown by solid lines (the fits for the 100.1 keV and 103.6 keV peaks are almost indistinguishable). The excluded peaks are shown by dashed lines.

Order of magnitude stronger limits were set by analyzing the 2nd stage data: $T_{1/2}$($^{184}$Os → 103.6) ≥ 8.9 × $10^{15}$ yr and $T_{1/2}$($^{186}$Os → 100.1) ≥ 4.4 × $10^{17}$ yr. Taking into account systematic uncertainties due to the uncertainties of the detection efficiencies, the intervals of fit and the isotopic compositions we obtained slightly reduced limits: $T_{1/2}$ ($^{184}$Os → 103.6) ≥ 6.8 × $10^{15}$ yr and $T_{1/2}$ ($^{186}$Os → 100.1) ≥ 3.3 × $10^{17}$ yr. The details of the 2nd stage data analysis were reported in [16]. The obtained limits exceed the present theoretical estimates of the decays' half-lives substantially (see Table 1) which motivated to carry out the 3rd stage of the measurement with a part of the osmium sample inside the cryostat of an HPGe detector.

The analysis of the data taken over 23840 h of the 3rd stage was performed in a similar way as for the two previous stages. Surprisingly there are negative peculiarities at both energies of interest: $S^{103.6}$ = (−108 ± 55) counts and $S^{100.1}$ = (−100 ± 55) counts, which resulted in rather small values of the excluded peaks areas: lim $S^{103.6}$ = 22.6 counts and $S^{100.1}$ = 24.6 counts. The result of the spectrum fits and excluded peaks are presented in Fig. 11. Taking into account the number of nuclei in the sample and the detection efficiency (see Table 3), one can get the following half-life limits, rather similar to the ones obtained in the 2nd stage of the measurement: $T_{1/2}$ ($^{184}$Os → 103.6) ≥ 8.2 × $10^{15}$ yr, $T_{1/2}$ ($^{186}$Os → 100.1) ≥ 5.5 × $10^{17}$ yr.

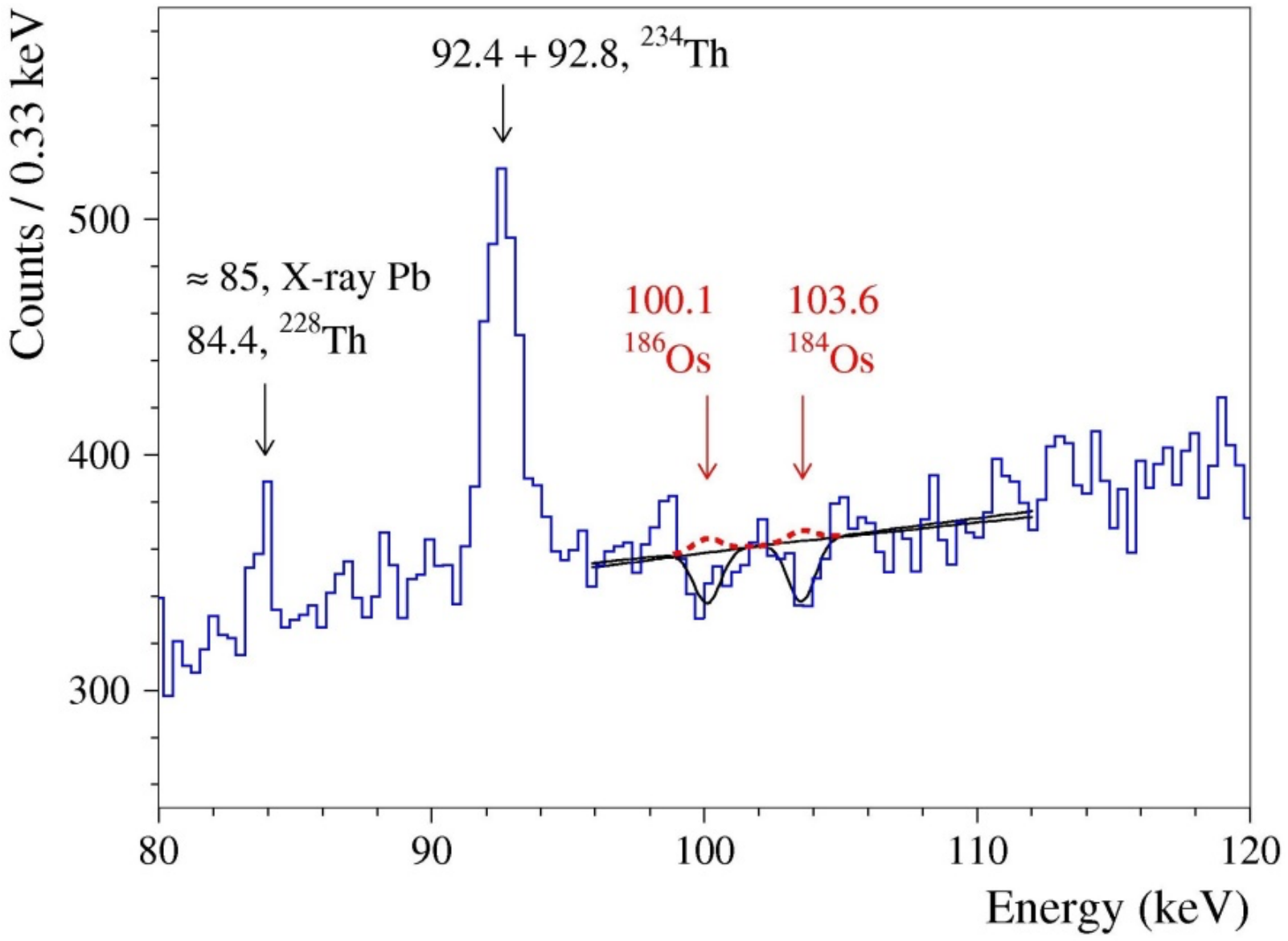

Fig. 11. Part of the energy spectrum measured for 23840 h in the 3rd stage of the measurement with the thin osmium sample placed directly on the Ge crystal of the HPGe detector *GS1*. The fits of the data are shown by solid lines. The excluded peaks with energies 100.1 keV and 103.6 keV are shown by red dashed lines.

The results obtained in the three stages of the measurement can be combined to increase the experimental sensitivity. At first, we calculate a half-life limit by using the obtained areas of the peaks with a slightly modified Eq. (4) (let's consider a general case of $n$ independent measurements):

$$\lim T_{1/2} = \frac{\ln 2 \cdot \eta \cdot \sum^{n} N_i \cdot \varepsilon_i \cdot t_i}{\lim S}, \quad (5)$$

where $N_i$, $\varepsilon_i$ and $t_i$ are numbers of nuclei of interest, the full absorption peak detection efficiencies for γ quanta and the times of measurements in the $i^{th}$ measurement, respectively. The value of lim $S$ can be obtained following the recommendations [62] using the peak areas $S_i$ and their standard deviations $\Delta S_i$ obtained in the fits of the individual measurements:

$$S = \sum^{n} S_i \pm \sqrt{\sum^{n} \Delta S_i^2}. \quad (6)$$

This approach, utilizing the data of the stages 2 and 3, gives half-life limits $T_{1/2}(^{184}\mathrm{Os} \rightarrow 103.6) \geq 1.8 \times 10^{16}$ yr and $T_{1/2}(^{186}\mathrm{Os} \rightarrow 100.1) \geq 1.0 \times 10^{18}$ yr. It should be noted that the use of the 1$^{st}$ stage in the analysis doesn't improve the limits due to a significantly lower detection efficiency of γ quanta with low energy emitted from the osmium ingots.

In the second approach a sum of the energy spectra taken in the 2$^{nd}$ and 3$^{rd}$ stages was analyzed. Again, adding the data measured in the 1$^{st}$ stage doesn't improve the experimental sensitivity due to the low detection efficiency. A part of the sum energy spectrum is shown in Fig. 12 together with the results of the fits by a model consisting in a linear function plus a peak with the energy expected in the α decays. The peaks were built from two Gaussians with the energy resolutions in the 2$^{nd}$ and 3$^{rd}$ stage, with areas determined by weight functions that depend on the number of nuclei of interest, the detection efficiency, and the time of measurement in the stages. The peaks areas obtained by the fits in the same energy interval as for the 1$^{st}$ and 2$^{nd}$ stage are $S^{103.6} = (-142 \pm 68)$ counts and $S^{100.1} = (-94 \pm 67)$ counts. The areas result in the excluded peaks areas: lim $S^{103.6} = 26$ counts and $S^{100.1} = 40$ counts. The excluded peaks are shown in Fig. 12 too.

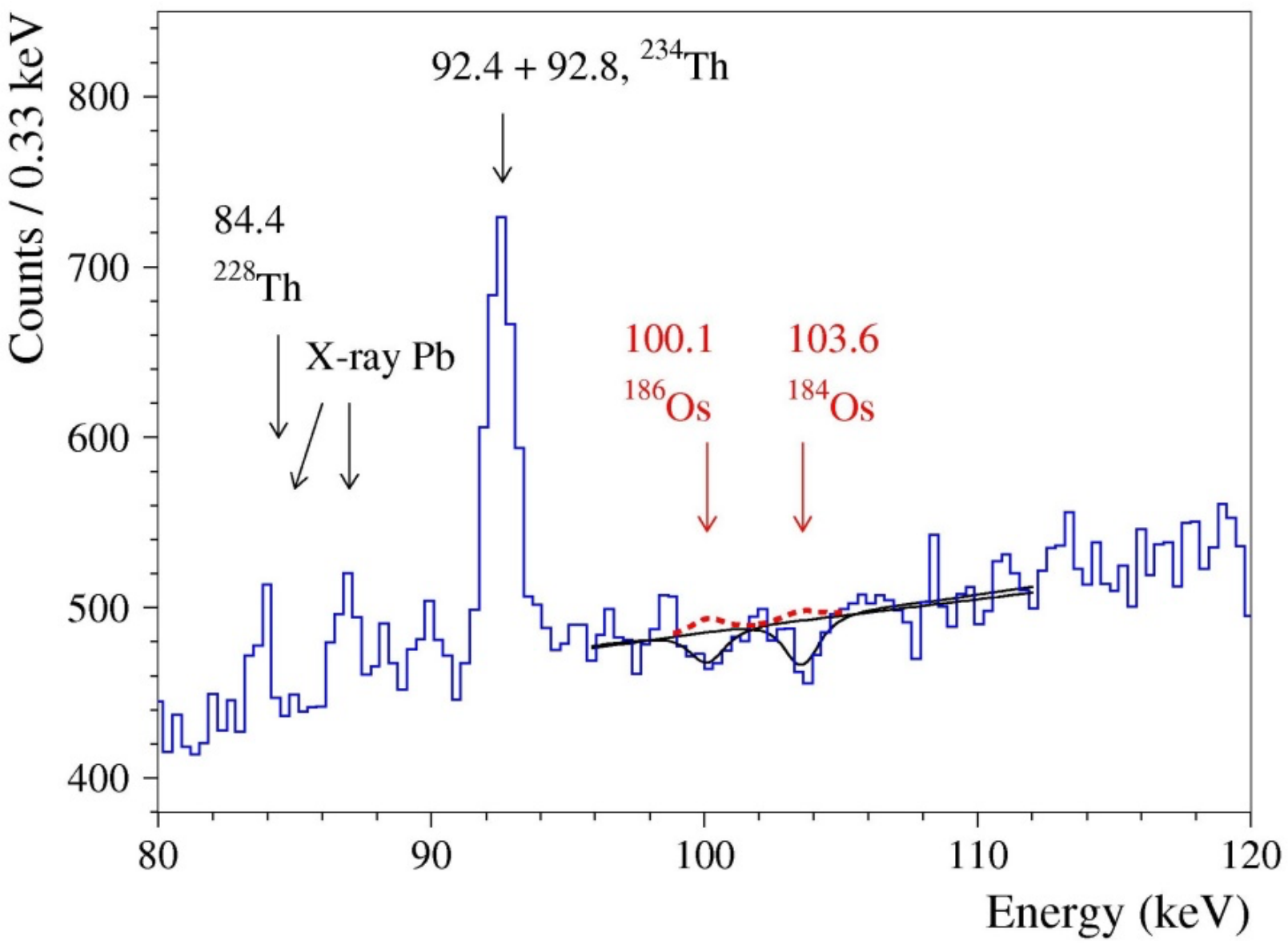

Fig. 12. Part of the sum energy spectrum measured in the 2nd and 3rd stages of the measurement (in total for 39691 h) with the osmium samples. The fits of the data are shown by solid lines, they are almost indistinguishable. The excluded peaks at 100.1 keV and 103.6 keV are shown by red dashed lines.

With Eq. (5) the following half-life limits were obtained: $T_{1/2}(^{184}\text{Os} \to 103.6) \geq 1.7 \times 10^{16}$ yr and $T_{1/2}(^{186}\text{Os} \to 100.1) \geq 8.2 \times 10^{17}$ yr. Quite expectable, the limits are comparable to the ones set by using the first approach. The second approach was used in the further analysis of other potentially $\alpha$ active naturally occurring osmium nuclides. Systematic uncertainties due to the uncertainties of the isotopic composition, detection efficiency and interval of fit were estimated as it was done in our previous report [16]. However, the increase of exposure by summing of the 2nd and 3rd stages just slightly improved sensitivity only for the two $\alpha$ transitions of $^{184}$Os, for the $\alpha$ decay of $^{186}$Os to the $2^+$ 100.1-keV excited level of $^{182}$W, and for the $\alpha$ transition of $^{188}$Os to the $4^+$ 364.1-keV excited level of $^{184}$W. For other nuclei and channels the highest sensitivity was achieved in the 2nd stage of the measurement. The half-life limits on $\alpha$ decays of the naturally occurring osmium isotopes accompanied by emission of $\gamma$ quanta are summarized in Table 1.

The experimental half-life limit for the $\alpha$ transition of $^{184}$Os to the first excited level of the progeny exceeds the theoretical estimations by a factor ~ 3−15. Similarly, the experimental limit for the $\alpha$ transition of $^{186}$Os is 2−15 times higher than the theoretical estimates. The limits for other nuclides and channels of the decay are far below the theoretical predictions. A breakthrough in the experimental sensitivity to the decays of $^{186}$Os, and especially to $^{184}$Os, could be achieved using enriched isotopes. There exists a possibility to enrich the osmium isotopes by gas centrifugation.

### 3.2. Search for 2β decay of $^{184}Os$ and $^{192}Os$

In the experimental data there are no peculiarities which could be ascribed to 2β decay processes in $^{184}Os$ or $^{192}Os$ with emission of γ quanta. Different combinations of the data were analyzed to set half-life limits on the 2EC and EC$\beta^+$ decay processes in $^{184}Os$ and the $2\beta^-$ decay of $^{192}Os$ to the $2^+$ 316.5 keV excited level of $^{192}Pt$.

#### *3.2.1. 2EC and ECβ⁺ decay of ¹⁸⁴Os*

The highest sensitivity to the theoretically most probable process of double electron captures with neutrino emission (2ν2EC) in $^{184}Os$ at level of lim $T_{1/2}$ ~ $10^{16}$ yr was achieved in the 2nd stage of the measurement thanks to a rather high detection efficiency of the BEGe detector in the energy region of X-rays (58–69 keV) expected in the decays [54]. Unfortunately, the detection efficiency of the GS1 detector used in the 3rd stage, despite localization of the sample directly on the Ge crystal, is an order of magnitude lower due to a thick dead layer on top of the Ge crystal: 1.74 mm. Similarly, utilization of additional statistics accumulated in the 3rd stage also doesn't improve the limit for the 2EC transition of $^{184}Os$ to the first $2^+$ 111.2-keV and to the $2^+$ 903.3-keV excited levels of $^{184}W$ obtained in the 2nd stage [54].

However, an advantage of the bigger exposure and higher detection efficiencies for γ quanta of higher energies allowed improving most of the half-life limits for the 2EC and EC$\beta^+$ processes in $^{184}Os$, e.g., a higher sensitivity was achieved for the 2EC in $^{184}Os$ to the $0^+$ 1002.5 keV excited level of $^{184}W$ accompanied by emission of γ quanta with energy 891.3 keV since the detection efficiency at this energy is rather high in all the stages: $\varepsilon_1$ = 4.505%, $\varepsilon_2$ = 2.397% and $\varepsilon_3$ = 5.542%[2]. A sum energy spectrum of the 1st, 2nd and 3rd stage of the measurement was analyzed to estimate the presence of the 891.3-keV γ peak expected in the transition. The spectrum was fitted by a simple model consisting of a linear function and a Gaussian peak at 891.3 keV. The width of the peak was bounded within the standard deviation by using the dependence of energy resolution (FWHM) on energy determined by analyzing several peaks present in the sum data taken in total for 42432 h: 46.5 keV ($^{210}$Pb), 238.6 keV ($^{212}$Pb), 295.2 keV and 351.9 keV ($^{214}$Pb), 338.3 keV and 911.2 keV ($^{228}$Ac), 609.3 keV, 1120.3 keV and 1238.1 keV ($^{214}$Bi), 583.2 keV ($^{208}$Tl), 661.7 keV ($^{137}$Cs), 1173.2 keV and 1332.5 keV ($^{60}$Co), 1460.8 keV ($^{40}$K) as $R_{\gamma}^{123}(\mathrm{keV}) = \sqrt{0.73(6) + 0.00280(10) \times E_{\gamma}}$, where $E_\gamma$ is in keV. The fit returns the peak area $S$ = (−5.6 ± 11.6) counts that corresponds to an excluded peak area lim $S$ = 13.9 counts. The sum energy spectrum in the vicinity of the expected 891.3-keV γ-peak, the fitting curve and the excluded peak are shown in Fig. 13. A slightly modified formula for a half-life limit from several measurements, with the γ quanta emission intensity included in the full absorption peak detection efficiencies, was used for combined data analysis:

$$\lim T_{1/2} = \frac{\ln 2 \cdot \sum^{n} N_i \cdot \varepsilon_i \cdot t_i}{\lim S}. \qquad (7)$$

[2] Certainly, statistical fluctuations of data at certain energy could impair the experimental sensitivity even for higher energies, as it happened with the 2EC transition of $^{184}Os$ to the $2^+$ 903.3-keV excited level of $^{184}W$. In this case the value of lim $S$ for 903.3-keV peak turned out to be rather big in all the cases when the data of the 1st or (and) 3rd stages was added to the data of the 2nd stage.

Taking into account the number of $^{184}$Os nuclei in the samples used in the 1st, 2nd and 3rd stage (listed in Table 3) and the above given detection efficiencies, one can obtain the following limit for the transition:

$$\lim T_{1/2}\,(^{184}\mathrm{Os},\ 2\nu2\mathrm{EC} \rightarrow 0^{+}\ 1002.5) = 4.4 \times 10^{17}\ \mathrm{yr}. \qquad (8)$$

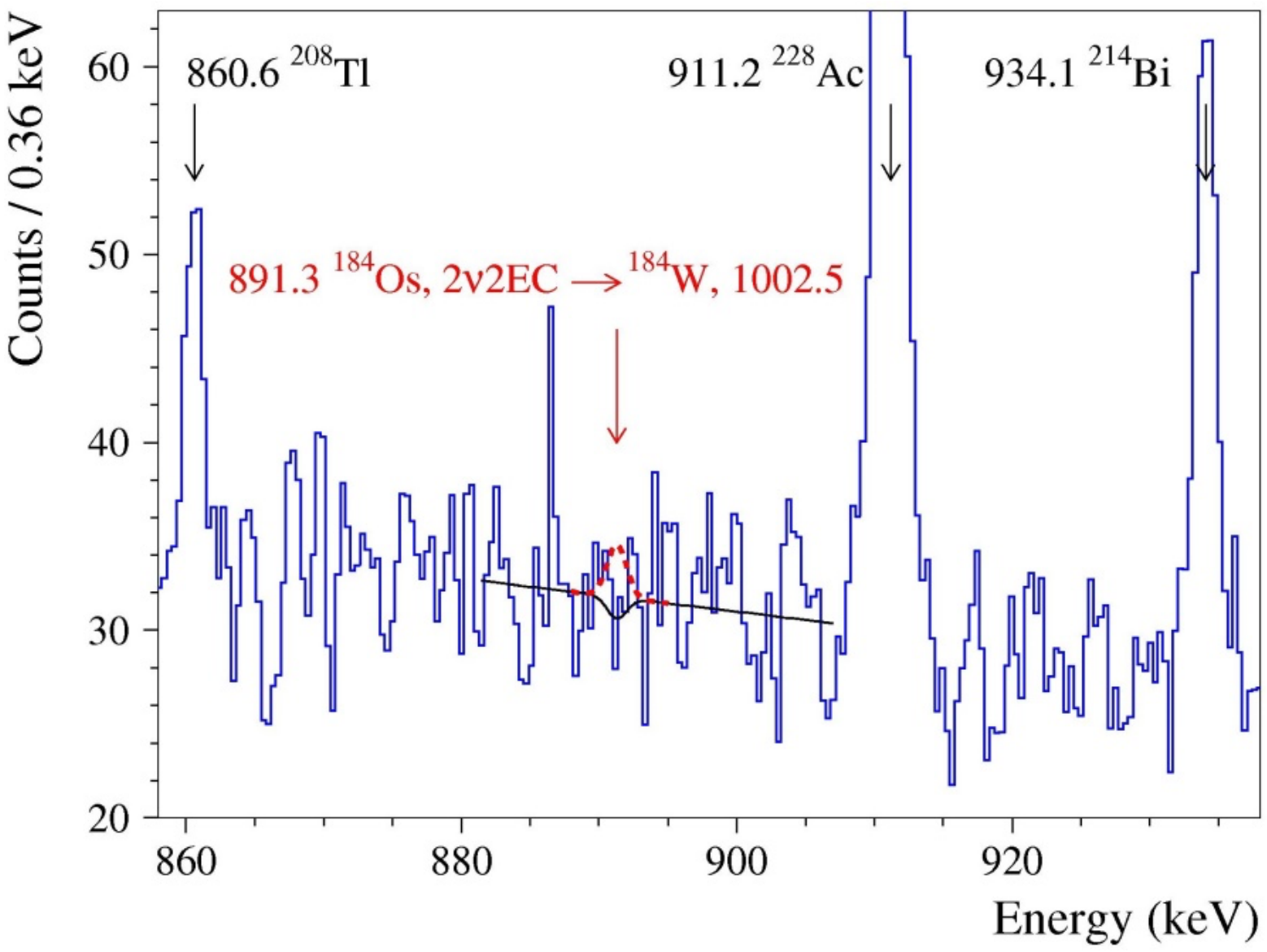

Fig. 13. The sum energy spectrum measured with the Os samples in the 1st, 2nd and 3rd stages (in total for 42432 h) in the energy region where an 891.3-keV γ peak after the 2$\nu$2EC decay of $^{184}$Os to the $0^+$ 1002.5-keV excited level of the progeny is expected. The fit of the data is shown by the solid line, while the excluded peak is shown by the dashed line. The γ peaks of $^{208}$Tl, $^{228}$Ac and $^{214}$Bi present in the data demonstrate a high spectrometric performance of the long-time measurement.

A limit on the neutrinoless mode of the decay is a bit lower:

$$\lim T_{1/2}\ (^{184}\mathrm{Os},\ 0\nu2\mathrm{EC} \rightarrow 0^+\ 1002.5) = 3.6 \times 10^{17}\ \mathrm{yr}, \quad (9)$$

since the detection efficiencies for this mode are smaller ($\varepsilon_1$ = 4.148%, $\varepsilon_2$ = 2.032% and $\varepsilon_3$ = 4.393%) due to an additional bremsstrahlung γ quantum emitted to release the energy carried away by two neutrinos in the 2ν mode.

Limits on other 2EC transitions of $^{184}$Os to excited levels of the progeny were obtained in a similar way by using different combinations of the experimental data. The results of the analysis are presented in Table 5. It should be noted that the best limit obtained in the 1st stage of the measurement for the 0ν2K decay of $^{184}$Os to the $2^+$ 111.2-keV excited level of $^{184}$W was corrected taking into account the measured isotopic concentration of the $^{184}$Os isotope ($\delta$ = 0.0170(7)% [16]) instead of the table value ($\delta$ = 0.02(1)% [63]) used in [56].

Table 5. Half-life limits on 2β processes in $^{184}Os$ and $^{192}Os$. $E_\gamma$ is the energy of γ (X-ray) quanta used in the analysis.

| Transition | Level of the progeny nucleus (keV) | $E_\gamma$ (keV) | Stage(s) of the measurement used in the analysis | Half-life limit, at 90% C.L. (yr) |
|---|---|---|---|---|
| **$^{184}Os \rightarrow {}^{184}W$** | | | | |
| 2ν2K | g.s. | 57–69 | 2 | $\geq 3.0 \times 10^{16}$ |
| 2νKL | g.s. | 57–69 | 2 | $\geq 2.0 \times 10^{16}$ |
| 2ν2K | $2^+$ 111.2 | 57–69 | 2 | $\geq 3.6 \times 10^{16}$ |
| 2νKL | $2^+$ 111.2 | 57–69 | 2 | $\geq 2.4 \times 10^{16}$ |
| 2ν2EC | $2^+$ 903.3 | 903.3 | 2 | $\geq 2.0 \times 10^{17}$ |
| 2ν2EC | $0^+$ 1002.5 | 891.3 | 1 + 2 + 3 | $\geq 4.4 \times 10^{17}$ |
| 2ν2EC | $2^+$ 1121.4 | 757.3 | 1 + 2 + 3 | $\geq 1.4 \times 10^{17}$ |
| 2νKL | $(0^+)$ 1322.2 | 903.3 | 2 | $\geq 1.7 \times 10^{17}$ |
| 2ν2L | $2^+$ 1386.3 | 1275.1 | 1 + 2 + 3 | $\geq 7.2 \times 10^{16}$ |
| 2ν2L | $(3)^+$ 1425.0 | 903.3 | 2 | $\geq 8.4 \times 10^{16}$ |
| 2ν2L | $2^+$ 1431.0 | 1319.8 | 1 + 2 + 3 | $\geq 8.3 \times 10^{16}$ |
| 0ν2K | g.s. | 1313.1–1314.5 | 1 + 2 | $\geq 2.0 \times 10^{17}$ |
| 0νKL | g.s. | 1370.4–1373.8 | 1 + 2 + 3 | $\geq 5.3 \times 10^{17}$ |
| 0ν2L | g.s. | 1427.9–1433.1 | 1 + 2 | $\geq 1.4 \times 10^{17}$ |
| 0ν2K | $2^+$ 111.2 | 1201.9–1203.3 | 1 | $\geq 3.1 \times 10^{17}$ |
| 0νKL | $2^+$ 111.2 | 57–69 | 2 | $\geq 1.9 \times 10^{16}$ |
| 0ν2EC | $2^+$ 903.3 | 903.3 | 2 | $\geq 1.7 \times 10^{17}$ |
| 0ν2EC | $0^+$ 1002.5 | 891.3 | 1 + 2 + 3 | $\geq 3.6 \times 10^{17}$ |
| 0ν2EC | $2^+$ 1121.4 | 757.3 | 1 + 2 + 3 | $\geq 1.3 \times 10^{17}$ |
| Near resonant 0νKL | $(0)^+$ 1322.2 | 903.3 | 2 | $\geq 1.7 \times 10^{17}$ |
| 0ν2L | $2^+$ 1386.3 | 1275.1 | 1 + 2 + 3 | $\geq 7.2 \times 10^{16}$ |
| 0ν2L | $(3)^+$ 1425.0 | 903.3 | 2 | $\geq 8.4 \times 10^{16}$ |
| Near resonant 0ν2L | $2^+$ 1431.0 | 1319.8 | 1 + 2 + 3 | $\geq 8.3 \times 10^{16}$ |
| 2νECβ$^+$ | g.s. | 511 | 1 + 2 + 3 | $\geq 2.6 \times 10^{17}$ |
| 2νECβ$^+$ | $2^+$ 111.2 | 511 | 1 + 2 + 3 | $\geq 2.6 \times 10^{17}$ |
| 0νECβ$^+$ | g.s. | 511 | 1 + 2 + 3 | $\geq 2.6 \times 10^{17}$ |
| 0νECβ$^+$ | $2^+$ 111.2 | 511 | 1 + 2 + 3 | $\geq 2.6 \times 10^{17}$ |
| **$^{192}Os \rightarrow {}^{192}W$** | | | | |
| 2β$^-$(2ν + 0ν) | $2^+$ 316.5 | 316.5 | 1 + 2 | $\geq 6.1 \times 10^{20}$ |

For the 0ν2EC capture in $^{184}$Os from K and L shells to the ground state of $^{184}$W one expects the energies of the bremsstrahlung γ quanta: $E_\gamma = Q_{2EC} - E_{b1} - E_{b2}$, where $E_{bi}$ are the binding energies of the captured electrons on K and L atomic shells of tungsten atom. Uncertainties of the energies $E_\gamma$ (arising from the uncertainty of the $Q_{2EC}$ value and of the difference between the $L_1$, $L_2$ and $L_3$ shells binding energies of tungsten atom) were taken into account in the analysis. The energy spectra were fitted by models consisting of a linear function and a Gaussian peak (with bound width using the dependences of the energy resolution on energy reported above). In the case of 0νKL decay also two nearby peaks of $^{214}$Bi were included in the model. The positions of the peaks searched for were varied inside the uncertainties and the biggest lim $S$ values obtained in the fits were accepted. The highest sensitivity to the 0ν2K and 0ν2L decays was achieved by analyzing the sum energy spectrum of the 1$^{st}$ and 2$^{nd}$ stage, while a sum of all stages was utilized to estimate the half-life limit for the 0νKL decay. The energy spectra used to derive limits on the 0ν2EC decays to the ground states of $^{184}$W, their fits and the excluded peaks are shown in Fig. 14. The obtained lower limits are given in Table 5.

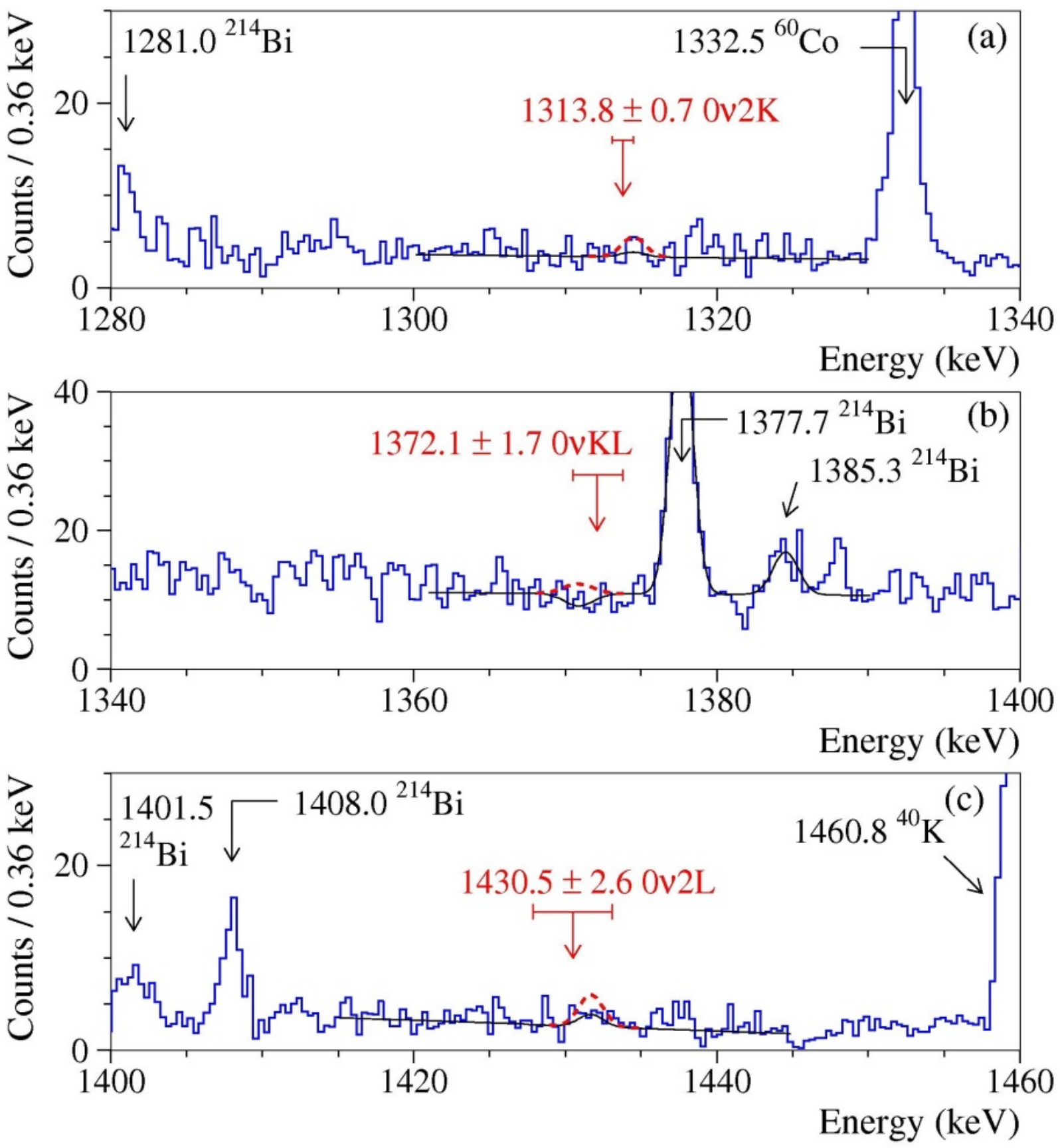

Fig. 14. Parts of the sum energy spectra measured with the Os samples in the 1st and 2nd stages (a, c) and in the 1st, 2nd and 3rd stage (b) where internal bremsstrahlung γ peaks from the 0ν2K (a), 0νKL (b), and 0ν2L (c) captures in $^{184}$Os to the ground state of $^{184}$W are expected. The horizontal lines (above the arrows labeling the energy of the peaks) show the energy bounds estimated from the uncertainty of the $Q_{2EC}$ value and the difference of the $L_1$, $L_2$ and $L_3$ shells binding energies of tungsten atom. Solid lines show the fits of the data, while the excluded γ peaks are represented by dashed lines. The biggest peak areas obtained in the fits were accepted for the estimation of the half-life limits.

The highest sensitivities to the EC$\beta^+$ decays to the ground state and to the $2^+$ 111.2 keV excited level of $^{184}$W were obtained by analyzing the sum energy spectrum of all stages. The half-life limits were calculated by using formulas Eq. 5 and Eq. 6. The limits for the 0ν and 2ν modes both for the transitions to the ground state and to the $2^+$ 111.2-keV excited level of the progeny are almost the same: $T_{1/2}$ ($^{184}$Os, EC$\beta^+$) $\geq 2.6 \times 10^{17}$ yr, due to rather similar detection efficiencies for the 511-keV annihilation γ quanta expected in all the decays. Corresponding diagrams for the 1st and 2nd stage can be found in the works [56] and [54], respectively, while the energy spectra gathered in the 3rd stage with the osmium sample and background measured without sample are shown in Fig. 15. There are (739 ± 47) counts in the annihilation peak in the spectrum taken with the sample, while in the background spectrum the 511-keV peak area is (216 ± 28) counts. After the normalization of the background data to the time of measurement with the osmium sample this leads to the residual peak area (−234 ± 135) counts. Using the Eq. 7 and the data of the 1st, 2nd, and 3rd stages of the experiment we got the following limit both for 2ν and 0ν modes of the EC$\beta^+$ decays to the ground state and to the $2^+$ 111.2 keV excited level of $^{184}$W:

$$\lim T_{1/2}\ (^{184}\text{Os, EC}\beta^+ \rightarrow \text{g.s., } 2^+\ 111.2) = 2.6 \times 10^{17}\ \text{yr}. \quad (10)$$

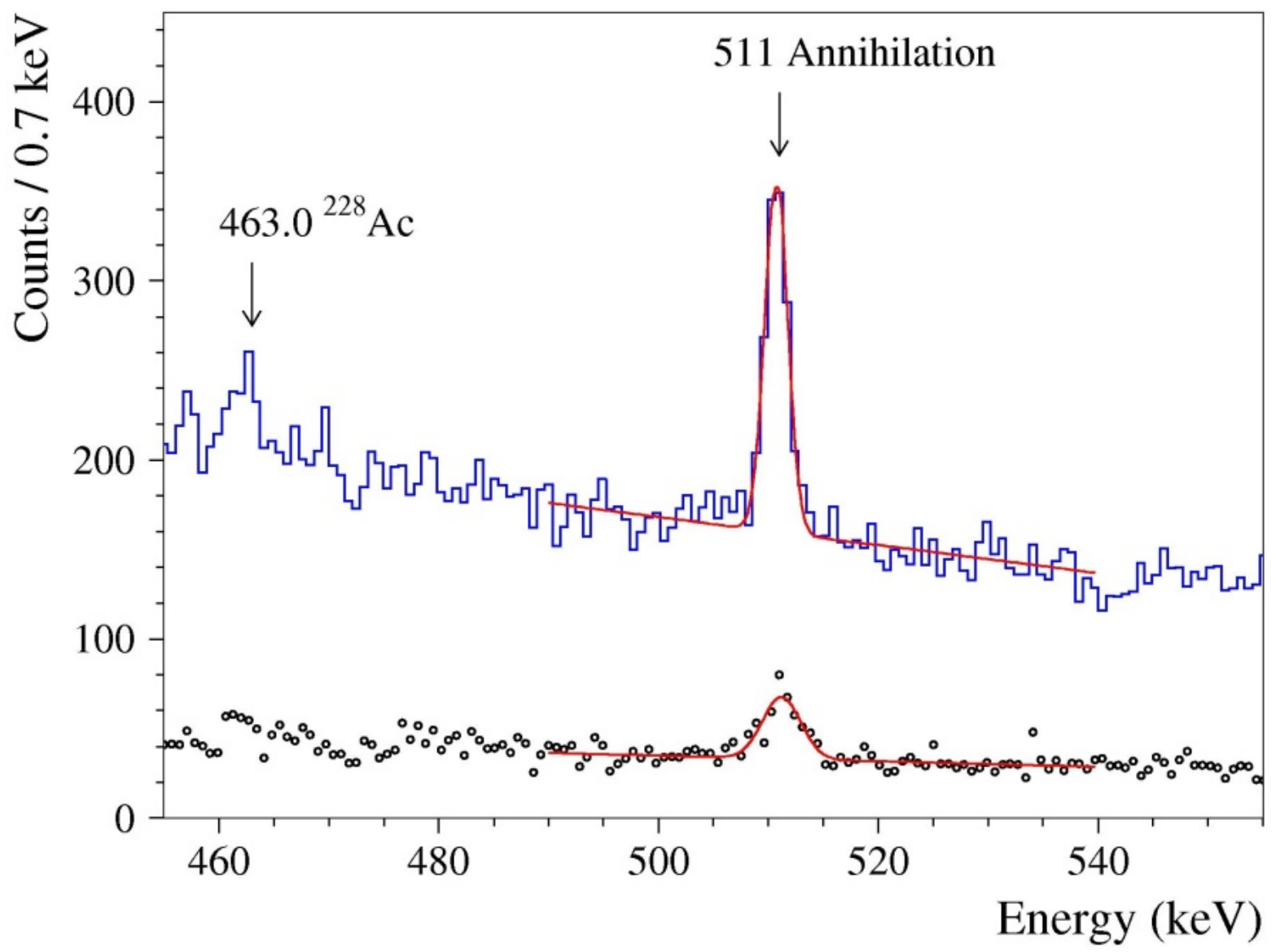

Fig. 15. Part of the energy spectrum measured with the Os sample in the 3rd stage of the measurement for 23840 h in the vicinity of the annihilation peak expected in the EC$\beta^+$ decay processes in $^{184}$Os both to the ground state and to the $2^+$ 111.2-keV excited level of $^{184}$W. The background data measured over 5292 h are presented as circles. The fits of the data are shown by red solid lines.

### *3.2.2. Search for 2$\beta^-$ decay of $^{192}$Os to the first $2^+$ 316.5 keV excited level of $^{192}$Pt*

The most sensitive search for the γ-peak with energy 316.5 keV, expected in the 2$\beta^-$ decay of $^{192}$Os to the $2^+$ 316.5 keV excited level of $^{192}$Pt, was achieved in the sum spectrum of the 1st and 2nd stages. To set a limit on the transition the sum energy spectrum was fitted by a linear function (to describe the continuous background) and a Gaussian centred at 316.5 keV with a bound width taking into account the dependence of the energy resolution on the energy obtained by analysing several γ peaks in the sum spectrum measured over 18592 h. The best fit was achieved in the energy interval 308−327 keV providing a peak area $S = 4.3 \pm 19.9$ counts, i.e. there is no evidence for the effect. Fig. 16 shows a part of the sum energy spectrum taken in the 1st and 2nd stage, the fitting curve and the excluded peak with an area of 36.9 counts. Taking into account the detection efficiencies for 316.5-keV γ quanta ($\varepsilon_1$ = 2.992% and $\varepsilon_2$ = 4.820%) the following limit was obtained:

$$\lim T_{1/2}\ (^{192}\mathrm{Os},\ 2\beta^- \rightarrow 316.5) = 6.1 \times 10^{20}\ \mathrm{yr}. \quad (11)$$

The limit is valid both for the 2ν and 0ν modes of the decay. All the half-life limits for the 2β processes in $^{184}$Os and $^{192}$Os are summarized in Table 5.

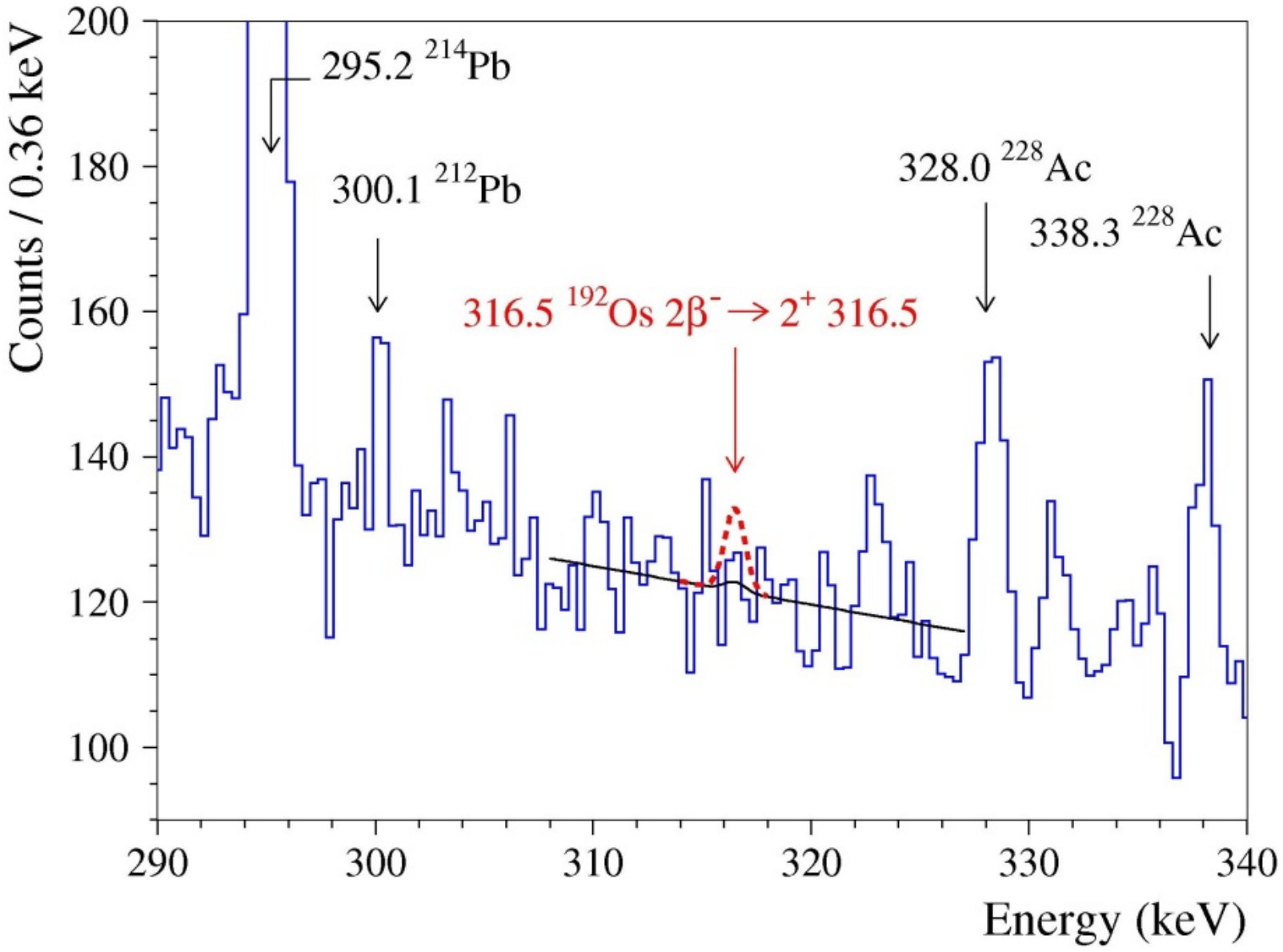

Fig. 16. Part of the sum energy spectrum measured with the Os samples in the 1st and 2nd stage over 18592 h in the energy region of the expected 316.5-keV γ peak from the 2β⁻ decay of $^{192}$Os to the first excited 2$^+$ 316.5 keV level of $^{192}$Pt. The fit of the data is represented by a black solid line, while the excluded γ peak is shown by a red dashed line.

All the limits obtained in the present work are very far from the theoretical predictions discussed in the Introduction section. Nevertheless, $^{184}$Os remains one of the most promising "double-beta plus" radionuclides due to its big atomic number (and thus comparatively large space factors and nuclear matrix elements) and to the possibility of near resonant 0ν2L captures to the (0)$^+$ 1322.2-keV and 2$^+$ 1431.0-keV levels of the progeny. The obstacle of its very low isotopic abundance can be overcome using the gas centrifugation method for isotopic enrichment, which would substantially improve the experimental sensitivity.

### 3.3. Search for 2α decay of the naturally occurring osmium isotopes

There are no peaks in the experimental data that can be attributed to 2α decay of Os isotopes with further emission of γ quanta by their progeny. Thus, only lower half-life limits were set by analyzing the different stages of the measurement.

In the case of 2α decay of $^{189}$Os (see a simplified decay scheme of $^{189}$Os in Fig. 4) one expects a γ peak at 482.2 keV originating from the β decay of $^{181}$Hf, progeny of $^{189}$Os. A fit of the sum spectrum taken with the osmium samples in the 1$^{st}$ and 2$^{nd}$ stage by a linear function (continuous background) and a Gaussian peak centered at 482.2 keV (the effect searched for) is shown in Fig. 17 (a). The fit returned a peak area $S = (-0.9 \pm 15.8)$ counts, which leads to lim $S$ = 25.1 counts. Taking into account the detection efficiencies for the γ quanta with energy 482.2 keV (that includes the γ quanta yield 80.5%), $\varepsilon_1 = 2.765\%$ and $\varepsilon_2 = 2.545\%$, the following half-life limit was set for the decay:

$$\lim T_{1/2}\,(^{189}\mathrm{Os}, 2\alpha) = 1.2 \times 10^{20}\ \mathrm{yr}. \quad (12)$$

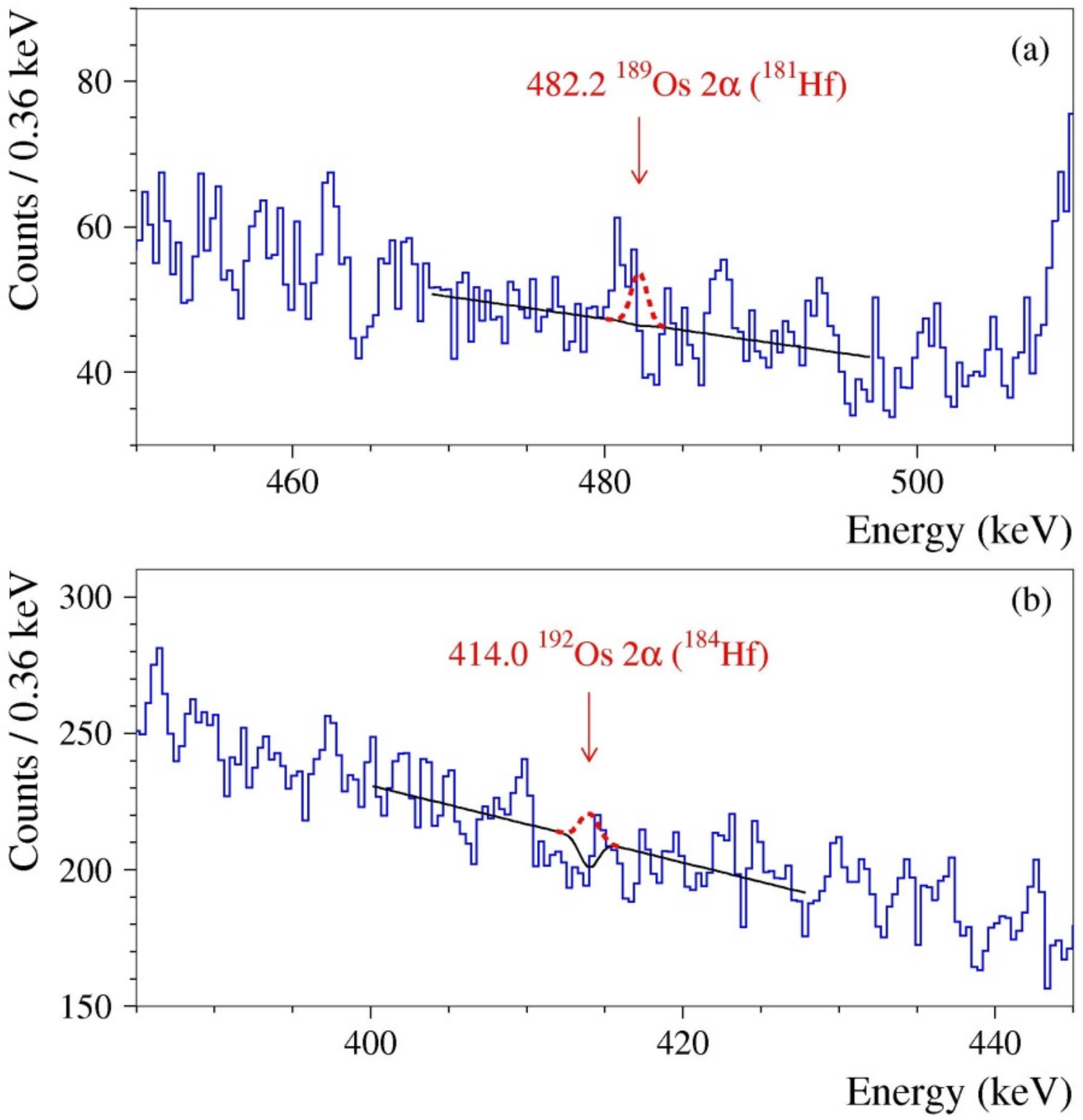

Fig. 17. Part of the sum energy spectrum measured with the Os sample in the 1st and 2nd stage over 18592 h in the region of the expected γ peak with energy 482.2 keV that is emitted in the β decay of $^{181}$Hf (product of 2α decay of $^{189}$Os) (a). Part of the sum energy spectrum of all the stages of the measurement over 42432 h in the vicinity of the expected 414.0-keV γ-peak of $^{184}$Hf progeny of the 2α decay of $^{192}$Os (b). The fits of the data by the background models are drawn by solid lines; excluded peaks are shown by red dashed lines.

The most stringent half-life limit for the 2α decay of $^{192}$Os was set analyzing the sum energy spectrum of all three stages of the measurement. In Fig. 17 (b) is shown the part of the spectrum in the vicinity of the expected 414.0-keV γ peak from the β decay of $^{184}$Ta, the progeny of $^{192}$Os after its 2α decay (see decay scheme in Fig. 5). The best fit using a Gaussian function (the effect searched for) and a linear function (continuous background) was achieved in the energy interval 400 – 428 keV, providing a peak area $S = (-42 \pm 45)$ counts, which corresponds to lim $S = 39$ counts, and, thus, to the following half-life limit:

$$\lim T_{1/2}\,(^{192}\mathrm{Os}, 2\alpha) = 3.1 \times 10^{20}\ \mathrm{yr}. \qquad (13)$$

Both limits represent the first attempts to search for 2α activity of naturally occurring osmium nuclides with emission of γ quanta. However, they are very far from the theoretical estimations (given in Table 2) and are mainly of methodological interest.

**4. Conclusions**

The search for rare decays (α, 2α and 2β) of naturally occurring osmium isotopes, accompanied by γ quanta emission, has been conducted over nearly 5 yr at the Gran Sasso underground laboratory of the INFN (Italy). Two semi-coaxial ultra-low background HPGe detectors and an ultra-low background BEGe detector were used to measure highly purified osmium samples of 99.999% purity with an initial mass of approximately 173 g (reduced to 118 and 59 g for the subsequent stages of the measurement). The primary objective of the measurement was to investigate the α decay of $^{184}$Os and $^{186}$Os to the first excited levels of their progeny, considering the promising theoretical estimations of their half-lives. These estimations were derived from several semi-empirical formulas: $T_{1/2} \sim (0.6 - 2.9) \times 10^{15}$ yr for $^{184}$Os and $T_{1/2} \sim (0.3 - 2.2) \times 10^{17}$ yr for $^{186}$Os. The final experimental limits set for the transitions, $T_{1/2} \geq 9.3 \times 10^{15}$ yr for $^{184}$Os and $T_{1/2} \geq 4.8 \times 10^{17}$ yr for $^{186}$Os, are substantially higher. However, it cannot be ruled out that the current theoretical models may not sufficiently describe the α transitions accurately with high precision. Further improvement of the experimental sensitivity could be achieved using enriched $^{184}$Os and $^{186}$Os samples. Measurements are in preparation and are expected to be at least 3−5 times more sensitive, which could be enough to detect the transitions.

As a by-product of the measurement, limits on different modes (two neutrino and neutrinoless) of double electron capture and electron capture with positron emission in $^{184}$Os to the ground and excited levels of $^{184}$W were determined to be in the range $T_{1/2} > 10^{16}$–$10^{17}$ yr. For the $2\beta^-$ decay of $^{192}$Os to the excited level $2^+$ 316.5 keV of $^{192}$Pt a limit $T_{1/2} \geq 6.1 \times 10^{20}$ yr was set. Despite the fact that all these limits are very far from the theoretical estimations, the investigation of $^{184}$Os is of certain interest, considering that this nucleus is amongst the most promising ones from a theoretical point of view. It has a big atomic number and the possibility of a near resonant 0ν2L capture to the $(0)^+$ 1322.2-keV and $2^+$ 1431.0-keV excited levels of the progeny. Furthermore, the experimental sensitivity could definitely be improved by isotopic enrichment using gas centrifugation.

For the first time limits on double-α decay of $^{189}$Os and $^{192}$Os were set at the level of lim $T_{1/2}$ ~$10^{20}$ yr. This was done by searching for the γ quanta of their comparatively short-lived β-active progeny. These limits are very far from the theoretical predictions, which are at level of $T_{1/2}$ ~ $10^{38}$–$10^{96}$ yr for $^{189}$Os and $T_{1/2}$ ~ $10^{148}$–$10^{301}$ yr for $^{192}$Os. Nevertheless, the obtained limits are a demonstration of the feasibility of ultra-low background γ-ray spectrometry to search for 2α decays with emission of γ quanta. One has also to keep in mind, that the theory of the 2α decay is under development and gives very different estimations depending on theoretical models applied in the calculations of the decay probability. The nuclide $^{184}$Os is the most interesting candidate to search for 2α decay considering the existence of reasonable theoretical estimations of its half-life at level of $T_{1/2}$ ~ $7 \times 10^{19}$ yr applying the symmetric 2α decay model. However, further investigation of this decay demands the use of experimental techniques that are sensitive to α particles, e.g., low temperature high energy resolution scintillating bolometric detectors.

**Acknowledgements**

F.A. Danevich, V.V. Kobychev and V.I. Tretyak were supported in part by the National Research Foundation of Ukraine Grant No. 2020.02/0011. D.V. Kasperovych was supported in part by the project "Investigation of rare nuclear and sub-nuclear processes" of the program of the National Academy of Sciences of Ukraine "Laboratory of young scientists". F.A. Danevich, O.G. Polischuk and V.I. Tretyak thank the INFN and the people of the DAMA group for the great support and kind hospitality in the difficult times during the Russian invasion of Ukraine.

**References**

---